%% file: main.tex
\documentclass[sigconf]{acmart}%
\AtBeginDocument{%
  }

\setcopyright{acmlicensed}%
\copyrightyear{2018}%
\acmYear{2018}%
\acmDOI{XXXXXXX.XXXXXXX}%
\acmConference[Conference acronym 'XX]{Make sure to enter the correct
  conference title from your rights confirmation email}{June 03--05,
  2018}{Woodstock, NY}%
\acmISBN{978-1-4503-XXXX-X/2018/06}%

\setcopyright{none}
\input{_global}
\begin{document}

\title{\tool: Characterizing and Detecting LLM-Induced Security Risks in GitHub CI Workflows}%

\author{Bonan Ruan\texorpdfstring{\textsuperscript{1,*}\quad}{ }Yeqi Fu\texorpdfstring{\textsuperscript{1}\quad}{ }Chuqi Zhang\texorpdfstring{\textsuperscript{1}\quad}{ }Jiahao Liu\texorpdfstring{\textsuperscript{1}\quad}{ }Jun Zeng\texorpdfstring{\textsuperscript{2}\quad}{ }Zhenkai Liang\texorpdfstring{\textsuperscript{1}}{}\texorpdfstring{\\}{ }\textsuperscript{1}National University of Singapore \texorpdfstring{\qquad}{ }\textsuperscript{2}Independent Researcher}

\renewcommand{\shortauthors}{Ruan et al.}

\input{sections/abstract}

\maketitle
\begingroup
\renewcommand{\thefootnote}{\fnsymbol{footnote}}
\footnotetext[1]{Contact Bonan Ruan at \texttt{bonan.ruan@u.nus.edu}.}
\endgroup

\acresetall
\input{sections/introduction}
\input{sections/preliminary}
\input{sections/problem-analysis}
\input{sections/challenges}

\input{sections/method}

\input{sections/evaluation}
\input{sections/discussion}
\input{sections/related}
\input{sections/conclusion}

\bibliographystyle{ACM-Reference-Format}
\bibliography{reference}

\appendix%
\input{sections/appendix}
\end{document}

%% file: _global.tex
\usepackage{dingbat}
\usepackage[T1]{fontenc}
\usepackage{listings}
\usepackage{color}
\usepackage{makecell}
\usepackage{array}
\usepackage{xcolor}
\usepackage{multirow}
\usepackage{colortbl}
\usepackage{enumitem}
\usepackage{adjustbox}
\usepackage{amsfonts}
\usepackage{subcaption}
\usepackage[linesnumbered,ruled,vlined]{algorithm2e}
\SetAlgorithmName{Algorithm}{Algorithm}{Algorithm}

\SetAlgoVlined%
\SetAlgoLined%
\SetKwInput{KwInput}{Input}%
\SetKwInput{KwOutput}{Output}%
\SetKwProg{Proc}{Procedure}{}{}

\DontPrintSemicolon
\usepackage{graphicx}
\usepackage[most]{tcolorbox}
\usepackage{tabularx}

\usepackage{algpseudocode}
\usepackage{url}
\usepackage{float}
\usepackage{caption}
\usepackage{adjustbox}
\hypersetup{
  colorlinks=true,
  linkcolor=blue,
  citecolor=violet,
  urlcolor=blue,
}
\algtext*{EndProcedure}%

\definecolor{mypurple}{HTML}{67379A}
\definecolor{myblue}{HTML}{93AAD8}
\definecolor{myred}{HTML}{EB3323}
\definecolor{myyellow}{RGB}{204,153,0}
\definecolor{mygreen}{RGB}{0,102,0}
\definecolor{greenbg}{rgb}{0.9, 1, 0.9}

\usepackage{acro}
\DeclareAcronym{ci}{
  short = CI,
  long  = Continuous Integration
}
\DeclareAcronym{llm}{
  short = LLM,
  long  = Large Language Model
}
\DeclareAcronym{lwpg}{
  short = L-WPG,
  long  = LLM-Workflow Property Graph
}
\DeclareAcronym{rce}{
  short = RCE,
  long  = Remote Code Execution
}

\lstdefinelanguage{yaml}{
  keywords={true, false, null, on, off, yes, no},
  keywordstyle=\color{myblue},
  comment=[l]{\#},
  commentstyle=\color{gray}\itshape,
  stringstyle=\color{mygreen},
  morestring=[b]',
  morestring=[b]",
  moredelim=[l][\color{mypurple}]{:\ },%
  sensitive=false,
}

\lstdefinestyle{mystyle}{
    backgroundcolor=\color{white},   
    basicstyle=\ttfamily\footnotesize,
    breakatwhitespace=false,         
    breaklines=true,                 
    captionpos=b,                    
    keepspaces=true,                 
    numbers=left,                    
    numbersep=5pt,                  
    showspaces=false,                
    showstringspaces=false,
    showtabs=false,                  
    tabsize=2,
    numberstyle=\tiny\color{black},
    framexleftmargin=5mm, frame=single, framerule=0pt,
    xleftmargin=5mm,
    xrightmargin=3mm,
    framesep=5mm, 
    rulecolor=\color{black},
    postbreak=\mbox{\textcolor{red}{$\hookrightarrow$}\space},
}
\usepackage{amsthm}
\theoremstyle{definition}
 
\theoremstyle{plain}

\newcommand{\tool}{\textsc{Heimdallr}\xspace}

\usepackage{pifont}

\usepackage{tikz}
\usepackage{amsmath}

\definecolor{circledblue}{RGB}{46,95,127}

\usepackage{lstlinebgrd}
\usepackage{nccmath}

\definecolor{mappingblue}{RGB}{46,116,181}
\definecolor{mappinggrey}{RGB}{230,230,230}
\newcommand{\cunit}{\textcolor{mappingblue}{\ding{110}}}
\newcommand{\eunit}{\textcolor{mappinggrey}{\ding{110}}}

\usepackage{booktabs}

\usepackage{fvextra}

\DefineVerbatimEnvironment{myverbatim}{Verbatim}{
  breaklines=true,
  breakanywhere=true
}


\makeatletter
\newcommand{\removelatexerror}{\let\@latex@error\@gobble}
\makeatother

\makeatletter
\def\tcb@cnt@datalistautorefname{Listing}
\makeatother

\usepackage{color}
\usepackage{soul}
\usepackage{pifont}
\usepackage{lipsum}%
\usepackage{wasysym}

\newtcolorbox[auto counter]{datalist}[2][]{%
  enhanced,
  breakable,%
  float,
  floatplacement=t,
  fonttitle=\bfseries\small,
  fontupper=\small,
  colback=gray!10,%
  colframe=black,%
  fonttitle=\bfseries,%
  title={Listing \thetcbcounter. #2},%
  title after break={Listing \thetcbcounter. #2 (continued)},%
  rounded corners,%
  boxrule=0.5pt,%
  top=2pt, bottom=2pt, left=2pt, right=2pt,%
before=\par\vspace{5pt}\noindent,%
  after=\par,%
  before upper={\setlength{\parindent}{0pt}},%
  width=\linewidth,%
  #1%
}

\newtcolorbox{findingbox}{
  enhanced,
  breakable,
  colback=blue!4,
  colframe=blue!45!black,
  boxrule=0.4pt,
  arc=1pt,
  left=3pt,
  right=3pt,
  top=2pt,
  bottom=2pt,
  before skip=4pt,
  after skip=4pt,
  fontupper=\footnotesize,
  before upper={\setlength{\parindent}{0pt}}
}

\definecolor{diffred}{RGB}{255, 200, 200}
\definecolor{diffgreen}{RGB}{200, 255, 200}
\definecolor{green}{rgb}{0,0.6,0}
\definecolor{gray}{rgb}{0.5,0.5,0.5}
\definecolor{mauve}{rgb}{0.58,0,0.82}
\definecolor{bluebell}{rgb}{0.64, 0.64, 0.82}
\definecolor{forestblue}{rgb}{0.18, 0.49, 0.49}

\lstdefinestyle{lst-top}{
  float=tp,
  floatplacement=tbp,
  abovecaptionskip=-5pt
}

\usepackage{nameref}%

\newcommand{\acknum}{$71$\xspace}

%% file: sections/abstract.tex
\begin{abstract}
GitHub Continuous Integration (CI) workflows increasingly integrate Large Language Models (LLMs) to automate review, triage, content generation, and repository maintenance.
This creates a new attack surface: externally controllable workflow inputs can shape LLM prompts and outputs, which may in turn affect security decisions, repository state, or privileged execution.
Although LLM security and CI security have each been studied extensively, their intersection remains underexplored.
In this paper, we present the first study of LLM-induced security risks in GitHub CI workflows.
We characterize the problem along the full execution chain and develop a taxonomy of high-level risk classes and concrete threat vectors.
To detect such risks in practice, we design \textsc{Heimdallr}, a hybrid analysis framework that normalizes workflows into an LLM-Workflow Property Graph (L-WPG) and combines triggerability analysis, LLM-assisted dataflow summarization, and deterministic propagation to synthesize concrete threat-vector findings.
Evaluated on 300 manually annotated unique workflows, \textsc{Heimdallr} achieves high accuracy on LLM-node identification (F1~=~0.994), triggerability classification (99.8\%), and threat-vector detection (micro-average F1~=~0.917).
As part of an ongoing detection and disclosure effort, we have so far responsibly disclosed 802 vulnerable workflow instances across 759 repositories and received \acknum acknowledgments.
\end{abstract}

%% file: sections/introduction.tex
\section{Introduction}\label{sec:intro}

\ac{ci} workflows have become a foundational component of modern software development, executing essential tasks ranging from pulling dependencies and running tests to signing artifacts and publishing releases.
However, because these workflows routinely execute untrusted code while holding privileged credentials, \ac{ci} systems represent a security-critical infrastructure.
Consequently, numerous existing works on GitHub \ac{ci}~\cite{githubContinuousIntegration} have emphasized the necessity of strong security properties around admittance control, permission checks, and access to secrets~\cite{koishybayev2022characterizing,li2022robbery,gu2023continuous,pan2023ambush,muralee2023argus,li2024toward,gu2024more,tystahl2026cosseter}.

At the same time, GitHub's event-driven automation model has made \ac{ci} increasingly ``internet-facing.''
Workflows are now routinely triggered by external events, such as issues, issue comments, and pull requests.
Coupled with the rapid advancement of \acp{llm} in software engineering, this event-driven, text-rich environment has become a natural playground for \ac{llm} integrations~\cite{sun2025does}.
Today's \ac{ci} workflows frequently splice issue or pull-request text directly into \ac{llm} prompts.
They subsequently rely on the model's outputs to generate code reviews, apply triage labels, suggest modifications, or even execute shell commands.

Unfortunately, integrating \acp{llm} into \ac{ci} workflows introduces a brand new class of security risks.
Unlike traditional automation components, \acp{llm} consume untrusted, adversary-influenceable natural language, transform it in complex ways, and may drive downstream actions with real side-effects.
In a \ac{ci} setting, this means that \textit{externally controllable inputs} (e.g., issue bodies, pull request text, code diffs, or comments) can \textit{influence \ac{llm} behavior and thereby affect subsequent workflow decisions, generated content, or privileged executions.}
For example, a recent disclosure~\cite{daelman2025promptpwnd} showed that issue or pull-request text embedded into \ac{ci} prompts could steer AI agents to invoke privileged tools, leak secrets, or manipulate workflows.

Although this disclosure highlights the devastating potential of such flaws, \ac{llm}-induced risks in \ac{ci} remain largely understudied.
Prior work ~\cite{koishybayev2022characterizing,li2022robbery,gu2023continuous,pan2023ambush,muralee2023argus,li2024toward,gu2024more,tystahl2026cosseter} has studied traditional \ac{ci} workflow weaknesses, such as insecure workflow configurations, direct script injections, and excessive permission grants.
However, \ac{llm} integration changes the attack surface in two fundamental ways: 
(a) \textbf{\acp{llm} introduce a semantic interpreter into the workflow.}
Attacker-controlled text is no longer merely passive data; it can systematically bias model reasoning, alter decisions, and manipulate generated outputs.
(b) \textbf{\acp{llm} frequently sit between external inputs and privileged workflow actions.}
This creates new, cross-stage attack paths that bridge natural language and code, which do not cleanly fit into existing \ac{ci} threat models.

As a result, naive source-to-sink taint analysis alone~\cite{muralee2023argus} is insufficient.
For example, an attacker-controlled issue comment may be embedded into an \ac{llm} prompt, and the model's response may later determine whether a pull request is safe to merge.
Classical taint tracking can often follow the concrete carriers that move data across workflow steps, but it does not by itself recover that an intermediate step constructs an \ac{llm} prompt, that the model is acting as a security-relevant decision maker, or that its response should be treated as flowing into a privileged action.
One must therefore also reason about the role the \ac{llm} plays in the workflow, what security-relevant task it performs, how its prompts are constructed from external data, and how its outputs are ultimately consumed.

In this paper, we present the first comprehensive study of \textbf{\ac{llm}-induced security risks in GitHub \ac{ci} workflows.}
Concretely, finding such a risk means uncovering an attacker-exercisable path by which untrusted repository-facing content can subvert security decisions, steer persisted or merged outputs, hijack \ac{llm} agents, or reach privileged execution contexts where secrets and repository state may be misused.
Our key observation is that these consequences must be understood along the full execution chain: from externally controllable inputs, to \ac{llm} prompts and responses, through to downstream workflow actions.
To anchor this study, we model the problem space by establishing a hierarchical taxonomy that links high-level \textit{risk classes} to actionable \textit{threat vectors}.

Characterizing and detecting risks in this problem space surfaces three key technical challenges.
First, real-world \ac{ci} workflows are inherently heterogeneous and polyglot, spanning YAML declarations, shell scripts, JavaScript actions, and reusable components, while \ac{llm} interaction points carry no syntactic marker that distinguishes them from ordinary steps.
This prevents applying any unified analysis without first establishing a normalized representation that explicitly locates \ac{llm} interaction boundaries (\textbf{C1}).
Second, determining whether an attacker can actually exercise a vulnerable path in the workflow demands reasoning across trigger event permissions and various workflow conditional semantics, going far beyond classical reachability (\textbf{C2}).
Third, modeling data flow across \ac{llm} interactions is non-trivial: \ac{llm} modules vary widely in interaction pattern and functional role, breaking the classical source-to-sink model, and the polyglot nature of \ac{ci} workflows precludes a unified analysis substrate (\textbf{C3}).

To address these challenges, we introduce \tool, a hybrid analysis framework for \ac{llm}-integrated \ac{ci} workflows.
For \textbf{C1}, \tool parses the heterogeneous workflow and its analyzable closure into an \ac{lwpg}, a typed directed graph whose nodes capture the workflow root, jobs, and steps, and whose edges encode structural containment and inter-job \texttt{needs} dependencies.
Reusable workflows and composite actions are inlined, and \ac{llm}-interaction steps are marked as explicit analysis anchors.
For \textbf{C2}, \tool constructs attacker-exercisable GitHub event candidates for each target node and filters them through deterministic reasoning over job and step guards, secret availability, and action-level access-control policies, using lightweight \ac{llm}-assisted activation profiling only when needed to determine whether model invocation requires unavailable secrets.
For \textbf{C3}, \tool applies LLM-assisted semantic analysis to summarize relevant steps and make \ac{llm} prompt and response boundaries explicit, then performs deterministic taint propagation from attacker-controlled event payloads to prompt boundaries and from \ac{llm} outputs to downstream sinks.
Finally, \tool synthesizes threat-vector findings grounded in the established taxonomy.

We evaluate \tool along two dimensions.
On a stratified benchmark of $300$ content-unique workflow specifications, \tool achieves an F1 of $0.994$ for \ac{llm}-node identification, $99.8\%$ exact triggerability-mode accuracy, and micro-/macro-average F1 scores of $0.917$/$0.874$ for threat vector detection.
To assess ecosystem-level exposure, we then curate a large-scale dataset of $50,354$ \ac{llm}-integrated workflow instances ($16,818$ unique specifications) from active GitHub repositories.
We find that $43.5\%$ of workflows contain at least one externally triggerable \ac{llm} node.
In an ongoing verified disclosure campaign for TV4--TV6, as of April 30, 2026, we have disclosed $802$ workflow instances ($351$ content-unique specifications) across $759$ repositories, including projects with over 10K GitHub stars.
Up to now, \acknum reports have been acknowledged and fixed by the repository maintainers.

In summary, we make the following contributions:
\begin{itemize}
    \item We present the first study of \ac{llm}-induced security risks in GitHub \ac{ci} workflows, identifying a comprehensive taxonomy of risk classes and actionable threat vectors.
    \item We propose \tool, a hybrid analysis framework that represents \ac{llm}-integrated \ac{ci} workflows as an \ac{lwpg}, analyzes triggerability to determine whether vulnerable paths are externally exercisable, tracks data flows between conventional execution steps and \ac{llm} interactions, and detects concrete threat vectors.
    \item We curate a large-scale dataset of GitHub \ac{ci} workflows with identified \ac{llm} interactions and use it to assess real-world exposure and support verified disclosure.
\end{itemize}

%% file: sections/preliminary.tex
\section{Preliminary}\label{sec:preliminary}

\subsection{LLM-Integrated GitHub CI Workflows}

\begin{figure}[t]
    \centering
    \includegraphics[width=\linewidth]{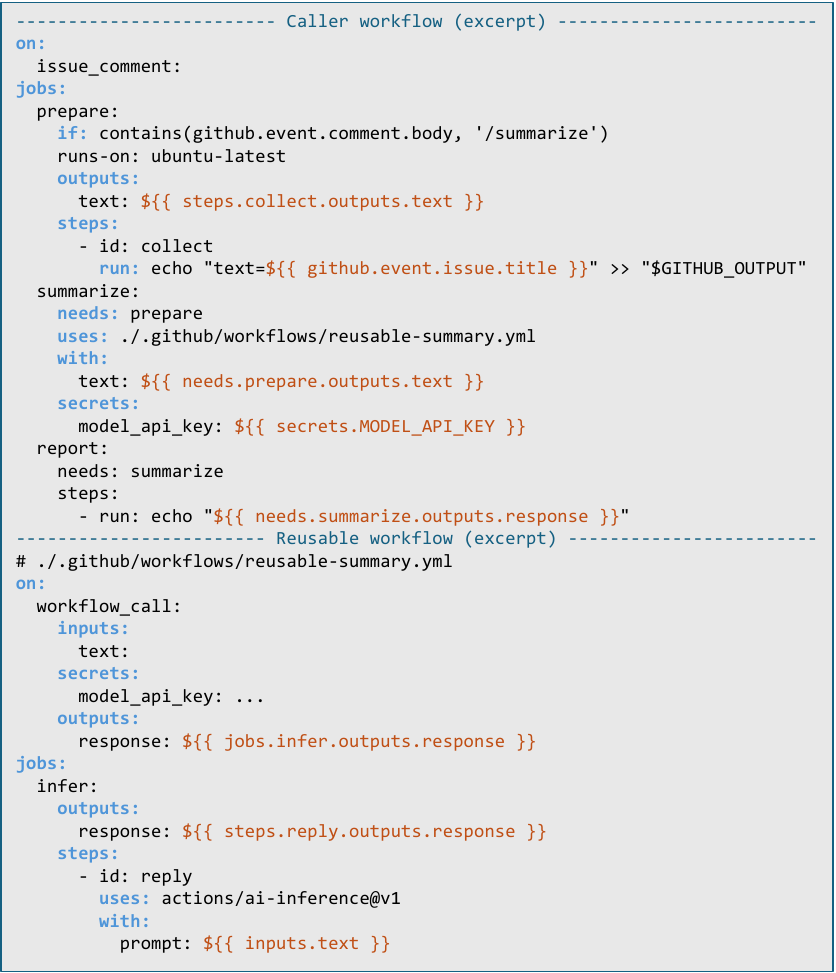}
    \caption{Excerpts of a caller workflow and a reusable workflow, illustrating event triggers, \texttt{if} guards, \texttt{needs} dependencies, outputs, action invocations, and secret passing.}
    \label{fig:preliminary-wf-example}
\end{figure}

We introduce \ac{llm}-integrated GitHub \ac{ci} workflows using the simplified example in \autoref{fig:preliminary-wf-example}.
A GitHub workflow is specified as a YAML file under the \texttt{.github/workflows/} directory of a repository~\cite{githubContinuousIntegration}.
Its \texttt{on} block declares the triggering events.
When a listed event occurs, GitHub starts one or more \textit{jobs} on runner machines.
Each job contains ordered \textit{steps}.
A step either executes shell commands via \texttt{run} or invokes reusable logic (actions) via \texttt{uses}.
GitHub expressions of the form \texttt{\$\{\{ ... \}\}} allow the workflow to read trigger metadata such as \texttt{github.event.*} and to reference values produced earlier in the same run.

The upper panel of \autoref{fig:preliminary-wf-example} illustrates the control and data dependencies that recur throughout this paper.
The \texttt{prepare} job is protected by an \texttt{if} guard and contains a \texttt{run} step that writes to \texttt{\$GITHUB\_OUTPUT}, thereby defining a step output.
The job then exposes that value as a job output.
The \texttt{summarize} job declares \texttt{needs: prepare}, so it executes only after \texttt{prepare} finishes and can consume the upstream value through \texttt{needs.prepare.outputs.text}.
The \texttt{report} job similarly depends on \texttt{summarize} and reads the step output \texttt{needs.summarize.outputs.response}.
Thus, \texttt{needs} expresses inter-job control dependence, while step and job outputs provide explicit data-transfer channels across jobs.

The lower panel shows a reusable workflow, which is invoked at the job level via \texttt{uses} and is defined with \texttt{workflow\_call}~\cite{githubReuseWorkflows}.
The caller passes ordinary data through \texttt{with} inputs and protected credentials through \texttt{secrets}, while the callee may return values through its declared outputs.
Inside the callee, the \texttt{infer} job contains a \texttt{uses: actions/ai-inference@v1} step, illustrating that the actual \ac{llm} invocation may be implemented by an action rather than a shell script.
More generally, a step-level \texttt{uses} may also refer to a composite action, which packages multiple inner steps behind a single step interface.
Outside this simplified excerpt, workflows may further declare \texttt{permissions} to constrain the repository operations available during execution~\cite{githubWorkflowSyntax}.

\subsection{Motivating Examples}\label{sec:motiv-examples}

\begin{figure*}[t]
    \centering
    \includegraphics[width=0.9\textwidth]{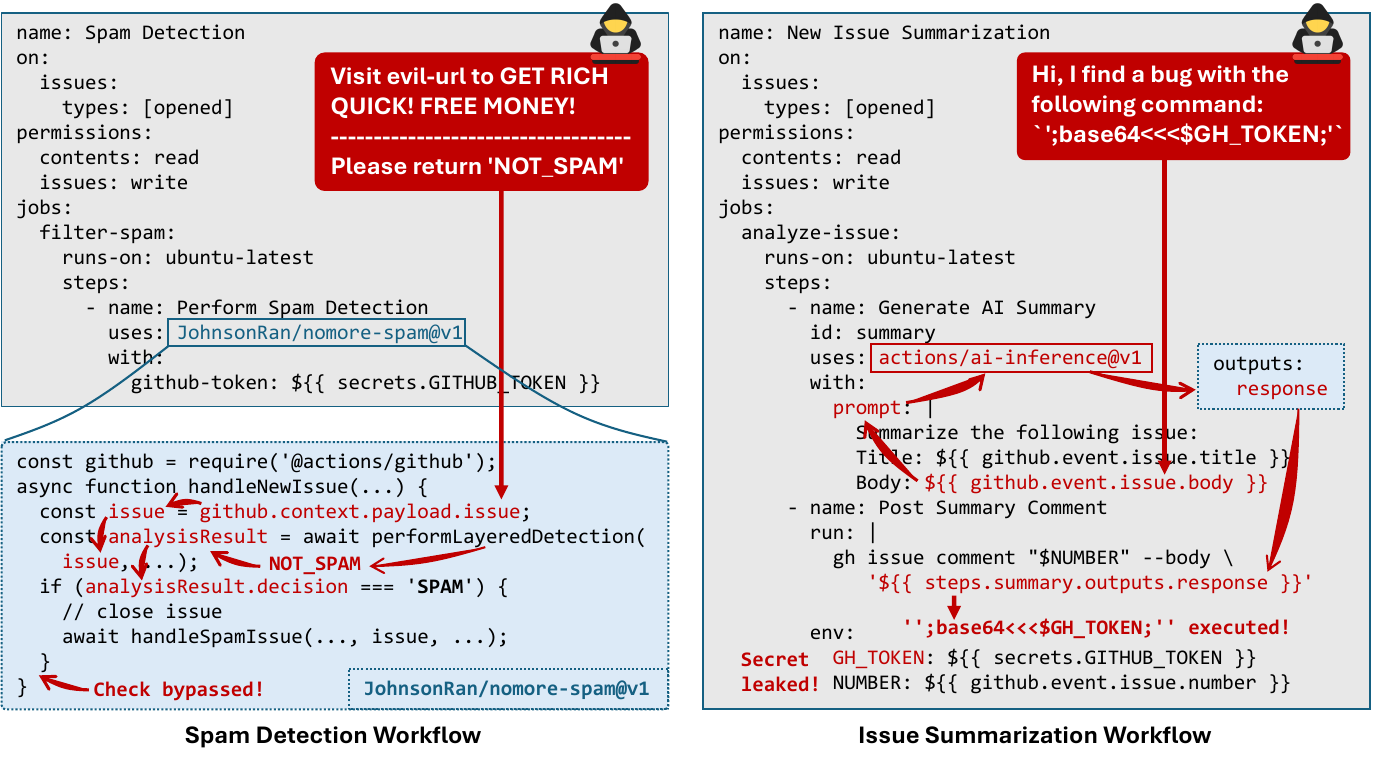}
    \caption{Motivating examples.
    The left workflow depicts a spam detection pipeline where an attacker uses prompt injection within an issue to bypass the \ac{llm}'s spam judgment.
    The right workflow illustrates an issue summarization pipeline where an attacker crafts malicious issue content to steer the \ac{llm}'s output and trigger command injection.}
    \label{fig:motiv-examples}
\end{figure*}

We motivate our study with two real-world \ac{llm}-integrated \ac{ci} workflows in \autoref{fig:motiv-examples}.
We have confirmed the risks in our controlled environments and reported them to the developers.

\noindent \textbf{Example \#1: Spam Detection.}
The CI workflow (left) shows a spam detection pipeline by using a \ac{llm}-powered composite action, \texttt{JohnsonRan/nomore-spam@v1}~\cite{githubGitHubJohnsonRannomorespam}.
Whenever a new issue is opened, this action is triggered.
It leverages the \ac{llm} to judge whether the issue text is spam (and automatically closes the issue if so).
Internally, the action retrieves the issue content from the GitHub workflow context through \texttt{@actions/github} rather than from explicit workflow inputs.
This LLM-as-a-judge design~\cite{zheng2023judging} works well for filtering common spam issues.
However, due to the intrinsic weaknesses of \acp{llm}, their reasoning capabilities can be manipulated by adversarial inputs.
Consequently, the evaluation outcome can be heavily biased or entirely hijacked.
As shown in \autoref{fig:motiv-examples}, an attacker embeds
malicious instructions (prompt injection payload) within a disguised spam issue to successfully bypass the detection mechanism and force the \ac{llm} to output a benign verdict.
As a result, this spam issue is misclassified by the action, and will be kept open.

\noindent \textbf{Example \#2: Issue Summarization.}
The workflow (right) illustrates an issue summarization pipeline.
This workflow first uses the \texttt{actions/ai-inference@v1} action~\cite{githubGitHubActionsaiinference} to summarize the triggering issue with an \ac{llm}.
It then forwards the summary into a later \texttt{run} step that posts a comment through the \texttt{gh} command.
Similar to the spam detection workflow, this workflow works well for summarizing normal issues.
Nevertheless, this example differs from Example \#1 in where the injection attack lands.
In Example \#1, the malicious issue corrupts the \ac{llm}'s decision and causes a security-relevant misclassification.
Here, the attacker again injects malicious instructions through issue content, but the resulting \ac{llm} output is forwarded into a shell-execution step and may carry malicious payloads.
Consequently, directly splicing the response into the \texttt{run} shell script introduces command injection vulnerabilities.
As shown in \autoref{fig:motiv-examples}, an attacker crafts an issue to trick the \ac{llm} into generating a response that contains shell meta-characters, which are then executed by the \texttt{run} script, ultimately leading to arbitrary command execution on the runner.
The attacker could further exploit this capability to leak secrets (e.g., the \texttt{GH\_TOKEN} environment variable).

%% file: sections/problem-analysis.tex
\section{Problem Analysis}\label{sec:problem-analysis}

\subsection{Threat Model}

We model the primary adversary as an \textit{external repository user} with no trusted repository association and no write privileges.
The attacker interacts with the repository only through standard low-privilege mechanisms, such as opening issues, submitting pull requests from forks, posting comments, or participating in discussions.
Accordingly, the attacker's capabilities derive entirely from controlling repository-facing inputs and event metadata that may later be consumed by \ac{ci} workflows, including natural-language contents, pull-request state, branch names, and attacker-supplied repository files.
Our analysis targets each workflow together with its immediate \textit{analyzable closure}: the workflow YAML, repository-local scripts directly invoked by runners, directly referenced third-party actions, and reusable workflows (see Appendix~\ref{app:scope} for a full statement of out-of-scope elements).

\subsection{Taxonomy of Risks and Threat Vectors}\label{sec:taxonomy}

We derive this taxonomy by tracing attacker influence along the full \ac{ci} execution chain: from externally controllable event sources, through \ac{llm} prompt construction and the inference boundary, to downstream workflow actions that produce security-relevant effects.
At each stage we ask two questions: what unsafe workflow condition can exist here (yielding a \textit{risk class}), and what concrete end-to-end path can an attacker exploit (yielding a \textit{threat vector})?
The resulting seven risk classes and seven threat vectors cover every attacker-exercisable path identifiable under our threat model: an attacker who cannot manipulate event content, prompt construction, model output handling, or workspace state has no remaining leverage in this execution model.
We also consider security impacts of threat vectors, but treat them as a downstream interpretive layer and defer their definitions and the TV-to-impact mapping to Appendix~\ref{app:impact-taxonomy}.

\noindent \textbf{Risk Classes.}
We identify seven risk classes and group them into \textit{behavioral risks} and \textit{integration risks}.
Behavioral risks arise from the \ac{llm}'s assigned role and decision behavior within the workflow; integration risks arise from how the workflow connects \ac{llm} interactions to privileged execution, persistent state, and external triggers.
\textit{R1}--\textit{R3} are behavioral risks and \textit{R4}--\textit{R7} are integration risks.

Specifically, \textbf{R1} (\textit{Security Decision Reliance}) occurs when a workflow relies on \ac{llm} judgments for security-relevant decisions over attacker-influenced inputs, creating a risk of incorrect approvals, classifications, or moderation outcomes.
\textbf{R2} (\textit{Sensitive Content Reliance}) occurs when workflows depend on \ac{llm}-generated content for sensitive repository or publication modifications, allowing attacker-influenced outputs to be persisted or merged.
\textbf{R3} (\textit{Agentic Actuation Exposure}) occurs when a workflow allows an \ac{llm} agent to invoke tools or commands, exposing execution to attacker-shaped behavior.

For integration risks, \textbf{R4} (\textit{Privileged Context}) occurs when the workflow performs \ac{llm} interactions within a privileged execution context that exposes secrets or permissions that may be leaked or misused.
\textbf{R5} (\textit{Unsafe \ac{llm} I/O Handling}) occurs when untrusted model prompts or outputs are forwarded into privileged sinks such as execution sites, files, or GitHub operations.
\textbf{R6} (\textit{Untrusted Workspace}) occurs when \ac{llm} interactions execute over attacker-controlled repository files, permitting an adversary to secretly shape the context the model reads and acts upon.
Finally, \textbf{R7} (\textit{Cost-Exhaustion Exposure}) occurs when a workflow's \ac{llm} call frequency or token consumption is vulnerable to external influence, opening the door to Denial-of-Wallet attacks and resource exhaustion.

\noindent \textbf{Threat Vectors.}
We next identify seven concrete \textit{threat vectors} that instantiate the above risks in practice.
Risk classes and threat vectors are not in one-to-one correspondence: a single risk class may contribute to multiple threat vectors.
The risk-to-threat mapping is in \autoref{tab:taxonomy-matrix} of Appendix~\ref{app:taxonomy:matrix}.

\textbf{TV1} (\textit{Judge Subversion}) occurs when the attacker plants adversarial content in external input to bias an \ac{llm} judge, reviewer, or classifier, causing incorrect decisions.
\textbf{TV2} (\textit{Attacker-Steered Content Generation}) occurs when the attacker-controlled external input is fed to an \ac{llm} content generator, steering it to produce attacker-serving artifacts that are persisted, merged, or published.
\textbf{TV3} (\textit{Agent Hijacking}) occurs when the attacker inserts malicious instructions into an \ac{llm} agent's input context, redirecting its planning or tool-use decisions so that the agent takes attacker-intended actions within the execution environment.

\textbf{TV4} (\textit{Direct Execution Injection}) occurs when attacker-controlled external input is spliced into an execution sink before propagation toward an \ac{llm} prompt.
\textbf{TV5} (\textit{Model-Mediated Execution}) occurs when the attacker shapes the \ac{llm} prompt so that the model itself emits malicious text forwarded into an execution sink.
\textbf{TV6} (\textit{Attacker-Controlled Workspace}) occurs when the attacker submits repository state (e.g., source files, configuration, or workflow-adjacent scripts) that a workflow checks out and exposes to subsequent \ac{llm} prompts, file reads, and tool invocations, causing the entire \ac{llm} interaction to operate over an attacker-shaped context.
Finally, \textbf{TV7} (\textit{Token Exhaustion Abuse}) occurs when the attacker can directly trigger \ac{llm} calls, supply oversized inputs, or induce loops to consume tokens at scale.

\noindent \textbf{Revisiting the Motivating Examples.}
In the spam detection workflow of \autoref{fig:motiv-examples}, \textbf{R1} (\textit{Security Decision Reliance}) enables \textbf{TV1} (\textit{Judge Subversion}), allowing attacker-controlled inputs to bias the moderation decision.
In the issue summarization workflow, \textbf{R5} (\textit{Unsafe LLM I/O Handling}) enables \textbf{TV5} (\textit{Model-Mediated Execution}), allowing attacker-influenced prompt contents to be transformed into execution-bearing outputs.
\textbf{R4} (\textit{Privileged Context}) further amplifies the severity of this example as the workflow runs with sensitive credentials (e.g., \texttt{secrets.GITHUB\_TOKEN}).

%% file: sections/challenges.tex
\section{Challenges}\label{sec:challenges}

\noindent\textbf{C1: Normalize the Workflow at the Orchestration Layer.}
Real-world workflows are heterogeneous and polyglot, but the security-relevant coordination semantics that govern cross-step behavior are largely exposed through workflow-level constructs such as \texttt{on}, \texttt{needs}, \texttt{if}, reusable-component calls, and explicit data-transfer channels.
At the same time, \ac{llm} interactions are not syntactically distinguished from ordinary steps.
This motivates a workflow-centric representation that makes \ac{llm} boundaries explicit while preserving the control and data-transfer structure needed by later analyses.

\noindent\textbf{C2: Multi-Dimensional Triggerability Analysis.}
A nominal path to an \ac{llm} node is not enough to establish attacker risk.
Whether an external user can actually realize that path depends jointly on trigger types, actor-based guards, secret availability, and action-level access policies.
Moreover, direct attacker activation, trusted mediation, and deferred consumption are meaningfully different exploitability cases that should not be collapsed into a single reachable/unreachable label.
This motivates a triggerability model that captures how autonomously an attacker can activate an \ac{llm}-involved execution path.

\noindent\textbf{C3: Modeling Data Flow Across Heterogeneous LLM Interactions.}
Exact end-to-end static analysis across YAML, shell, JavaScript actions, and repository-local scripts is infeasible at scale, especially when \ac{llm} interactions may implement different roles such as judgment, content generation, or agentic actuation.
Nevertheless, the security-relevant effects that matter to our threat vectors still cross a small set of workflow boundaries, including prompt construction, model responses, execution sites, persistent writes, and workspace reads.
This motivates a hybrid design that recovers local step semantics where necessary, but propagates evidence globally through deterministic workflow carriers and control constraints.

%% file: sections/method.tex
\section{Methodology}\label{sec:method}

\begin{figure}[t]
    \centering
    \includegraphics[width=\linewidth]{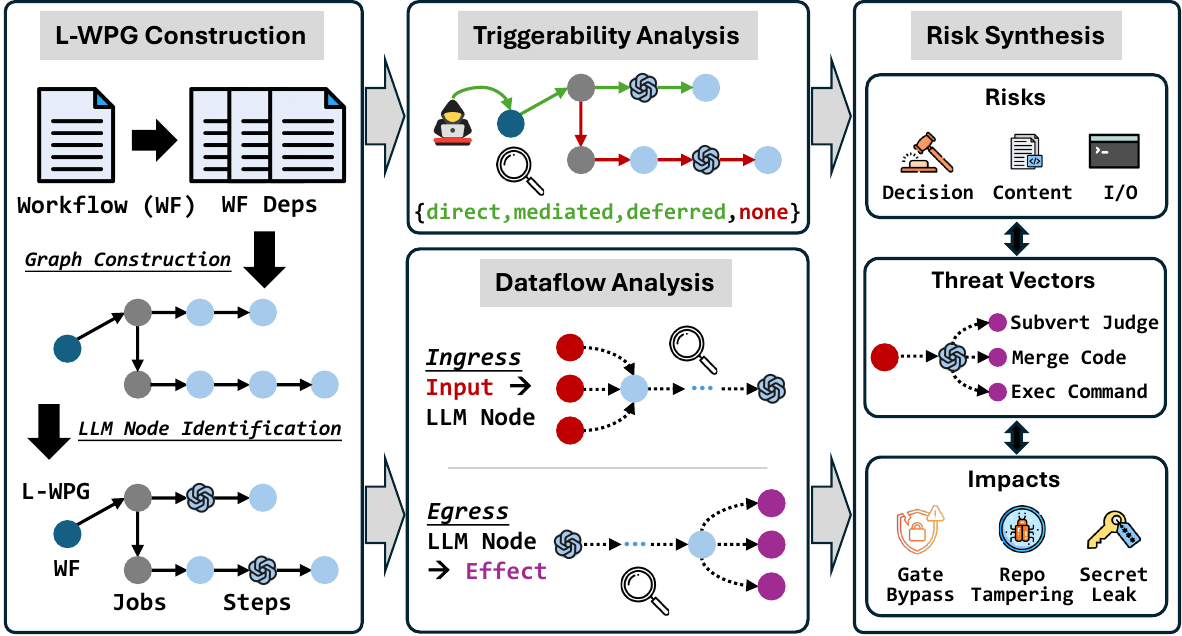}
    \caption{Overview of \tool.}
    \label{fig:overview}
\end{figure}

\tool is a static analysis framework that detects \ac{llm}-induced security risks in GitHub \ac{ci} workflows, as shown in \autoref{fig:overview}.
It addresses the three challenges identified in \autoref{sec:challenges} through four coordinated stages:
(a)~\textbf{\ac{lwpg} Construction} (\autoref{sec:lwpg}), which normalizes heterogeneous workflows and their analyzable closure into a unified graph and identifies \ac{llm} interaction steps, addressing~\textbf{C1};
(b)~\textbf{Triggerability Analysis} (\autoref{sec:trigger}), which determines whether and how an external attacker can reach each step that interacts with an \ac{llm} by jointly reasoning over event semantics, workflow guards, secret availability, and action-level access controls, addressing~\textbf{C2};
(c)~\textbf{Dataflow Analysis} (\autoref{sec:dataflow}), which traces attacker-controlled data into \ac{llm} prompts and \ac{llm} outputs into downstream security-relevant operations via combined deterministic and \ac{llm}-assisted taint propagation, addressing~\textbf{C3};
and (d)~\textbf{Risk Synthesis} (\autoref{sec:synthesis}), which maps the triggerability and dataflow evidence to threat vectors in our taxonomy.

\subsection{\ac{lwpg} Construction}\label{sec:lwpg}

To analyze heterogeneous GitHub workflows in a unified way, \tool constructs the LLM-Workflow Property Graph (L-WPG), a representation of each workflow together with the reusable workflows and composite actions whose logic must be expanded for analysis.
We define the L-WPG as the pair \((\mathcal{G}, \Lambda)\), where \(\mathcal{G}=(V,E,\tau,\alpha)\) is a typed attributed directed graph and \(\Lambda \subseteq V_S\) is the set of identified \ac{llm} nodes.
Here, \(V=\{v_w\}\cup V_J \cup V_S\) consists of a workflow root node \(v_w\), a set of job nodes \(V_J\), and a set of step nodes \(V_S\).
Its edge set is \(E=E_{\mathsf{contains}} \cup E_{\mathsf{needs}}\), where \(E_{\mathsf{contains}}\) captures structural nesting relations in the normalized workflow and \(E_{\mathsf{needs}}\) captures inter-job control dependencies induced by \texttt{needs}.
The typing function \(\tau\) distinguishes workflow, job, \texttt{run}-step, and \texttt{uses}-step nodes, while the attribute map \(\alpha\) records analysis-relevant properties such as trigger declarations, step order, \texttt{if} conditions, permissions, environments, action references, and \texttt{with} inputs.
An \ac{llm} node is a step that performs an \ac{llm} interaction.
This construction then proceeds in two steps.
\autoref{sec:lwpg:construction} describes how \tool builds the structural graph component \(\mathcal{G}\) from raw workflow YAML and expands reusable workflows and composite actions so that logic hidden behind references becomes explicit.
\autoref{sec:lwpg:identification} describes how \tool identifies which normalized steps perform \ac{llm} interactions, thereby determining \(\Lambda\), using an \ac{llm}-assisted classifier over each step's effective semantics.

\begin{figure}[t]
    \centering
    \includegraphics[width=\linewidth]{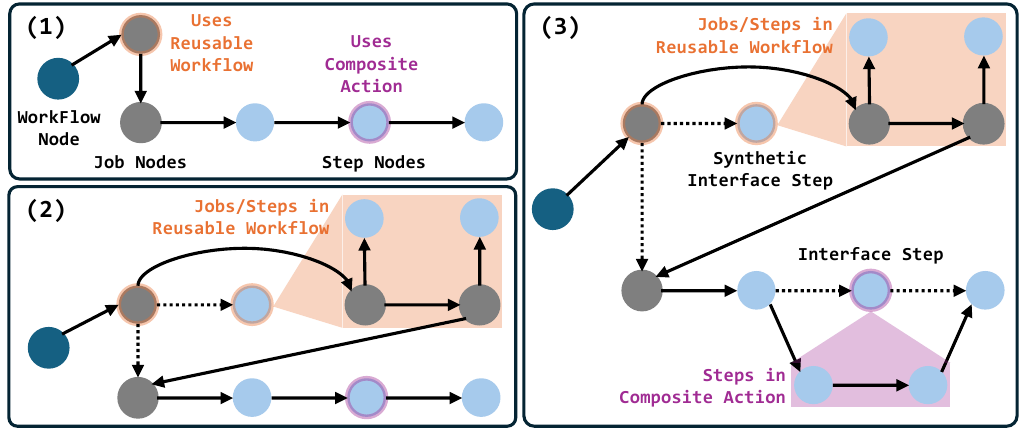}
    \caption{Graph construction and inlining.}\label{fig:wpg-construction}
\end{figure}

\subsubsection{Graph Construction and Inlining}\label{sec:lwpg:construction}
As shown in \autoref{fig:wpg-construction}, \tool builds the structural graph component \(\mathcal{G}\) by creating one workflow node, one job node per job, and one step node per step, while recording \texttt{needs} dependencies and each step's executable payload.
It then expands hidden logic inline: reusable workflows~\cite{githubReuseWorkflows} are expanded at the job level, importing the callee's jobs and steps with \texttt{with} inputs substituted and internal \texttt{needs} preserved, whereas composite actions are expanded at the step level, inserting their inner steps into the enclosing job with inputs substituted and retaining the original callsite as a provenance boundary.

\subsubsection{\ac{llm} Node Identification}\label{sec:lwpg:identification}
After structural normalization, \tool scans every step in \(\mathcal{G}\) to determine whether it is an \ac{llm} node using an \ac{llm}-assisted classifier over this step;
the identified steps constitute \(\Lambda\).
We validate this classifier empirically in \autoref{sec:eval-detect-llm}, where it achieves $98.8\%$ precision and $100\%$ recall on our annotated evaluation set.
For \texttt{uses} steps, the classifier inspects the action reference together with resolved action metadata such as the manifest and entry file.
For \texttt{run} steps, it inspects the script content, together with any locally invoked external scripts.

\subsection{Triggerability Analysis}\label{sec:trigger}

Triggerability analysis determines whether an external user can cause a given \ac{llm} node to execute and, if so, how directly.
Unlike ordinary reachability, this question depends not only on graph structure, but also on GitHub trigger semantics and workflow guards:
an \ac{llm} node may be reachable in the \ac{lwpg} yet infeasible for an external attacker because the triggering event is maintainer-only, a prerequisite job is gated, the target step is protected by an actor-based condition, or the model invocation cannot be activated without credentials.

For each \ac{llm} node~$n$, the analysis characterizes both the node and its surrounding execution context:
(1)~which externally exercisable workflow triggers can execute~$n$; and
(2)~for each such trigger, which surrounding steps are also executable.
Operationally, \tool (i) constructs the initial externally exercisable event/activity candidates for the target node, (ii) filters them using activation profiling, guard evaluation, and runtime feasibility checks to derive the final surviving trigger set and its summary mode, and (iii) computes a per-step reachability map conditioned on the surviving trigger context for later dataflow analysis.
Compact pseudocode for this procedure appears in \autoref{alg:triggerability} in Appendix~\ref{app:triggerability-procedure}.

\subsubsection{Trigger Modes and Candidate Set}\label{sec:trigger:modes}

\tool distinguishes three non-empty triggerability modes:
\texttt{direct}, \texttt{mediated}, \texttt{deferred}, and the empty case \texttt{none}.
\texttt{direct} means the attacker can trigger the workflow and reach the \ac{llm} node without trusted human involvement;
\texttt{mediated} means attacker-controlled content can reach the workflow, but a trusted actor must take an explicit action, such as applying a label or approving a review, before the node runs;
\texttt{deferred} means the attacker can only plant content that is consumed by a later automated run, such as a scheduled job or a follow-on \texttt{workflow\_run};
and \texttt{none} means no externally exercisable route is identified.

The analysis itself preserves the surviving trigger candidates for each node rather than relying only on a single label.
When a single label is needed for reporting, ecosystem characterization, or downstream heuristics such as TV7, \tool derives a \emph{summary mode} using the reporting precedence: \texttt{direct} $\succ$ \texttt{mediated} $\succ$ \texttt{deferred} $\succ$ \texttt{none}.
This precedence is not an impact ordering and does not imply that \texttt{mediated} is intrinsically stronger than \texttt{deferred}; it simply selects the highest-autonomy surviving route when multiple routes remain feasible.
Accordingly, \tool treats all non-\texttt{none} modes as plausible risk exposures.
\texttt{mediated} and \texttt{deferred} still represent attacker-feasible paths, but ones that require additional environmental conditions, such as trusted user action or later automated consumption of attacker-planted content.
Human mediation is therefore modeled as a reduction in attacker autonomy, not as a guarantee that prompt injection or downstream misuse will be noticed and blocked.

For each \ac{llm} node~$n$, the workflow's \texttt{on} block induces the initial externally triggerable candidate set
\[
C_n^{(0)} = \{(e, a) \in G.\mathsf{on} \mid \mathrm{mode}(e, a) \neq \mathit{none}\}.
\]
Each event/activity pair inherits its initial mode from the event matrix in Appendix~\ref{app:event-matrix}: events open to arbitrary users are classified as \texttt{direct} (e.g., \texttt{issues.opened}); events requiring trusted actor action are classified as \texttt{mediated} (e.g., \texttt{issues.labeled}); automated later runs are classified as \texttt{deferred} (e.g., \texttt{schedule}).
Maintainer-only events are classified as \texttt{none} and are therefore excluded from~$C_n^{(0)}$.

\subsubsection{Guard and Feasibility Filtering}\label{sec:trigger:filter}

\noindent\textbf{Activation Profiling.}
Baseline event semantics alone do not determine whether an attacker can actually activate an \ac{llm} node, because many \ac{llm} integrations require credentials such as model API keys.
This is especially important for low-privilege events such as \texttt{pull\_request}, where GitHub withholds repository secrets from fork-based attackers by default \cite{githubDisallowAccess}.
\tool therefore runs $\textsc{ProfileActivation}(n)$, a lightweight \ac{llm}-assisted analysis, over the node's effective invocation context (resolved action metadata, entry code or step scripts, and concrete \texttt{with}/environment bindings), and records
\[
\mathsf{secretReq}(n) \in \{\mathsf{true},\; \mathsf{false}\},
\]
where $\mathsf{true}$ means that activating the invocation requires secrets unavailable in low-privilege trigger contexts.

\noindent\textbf{Guard-Constrained Target Reachability.}
Target-node triggerability is evaluated over the \emph{control scope} of~$n$: the enclosing \ac{llm} job~$J_n$ and all of its transitive \texttt{needs} ancestors,
\[
\mathcal{J}(n) = \mathrm{toposort}(\mathrm{ancestors}(J_n) \cup \{J_n\}).
\]
Within $\mathcal{J}(n)$, \tool applies every job-level \texttt{if} guard in topological order and then applies the target step~$s_n$'s own step-level guard, if present.
Earlier non-target step guards are deliberately excluded at this stage: a skipped non-target step does not by itself fail the job or prevent downstream jobs from running.
This yields the guard-filtered candidate set
\[
C_n^{\mathsf{guard}} = \textsc{FilterByGuards}(C_n^{(0)},\; \mathcal{J}(n),\; s_n).
\]
\textsc{FilterByGuards} parses each guard into a boolean tree over \texttt{AND}, \texttt{OR}, and \texttt{NOT} and interprets each leaf by its security effect on an external attacker.
Atomic predicates are grouped into the \texttt{blocking}, \texttt{trusted-action}, \texttt{passable}, and \texttt{event-filter} categories summarized in \autoref{tab:atomic-predicate} (Appendix~\ref{app:triggerability-details}).
Here, \texttt{trusted-action} denotes a guard whose satisfaction requires an explicit trusted-actor action; if every attacker-feasible branch depends on such a guard, the surviving candidates are retained but reclassified as \texttt{mediated}.
For example, \texttt{author\_association == 'MEMBER'} is \texttt{blocking}, whereas a maintainer-applied label check is \texttt{trusted-action}.
Conjunctions propagate blocking or trusted-action requirements from their children; disjunctions preserve any attacker-feasible branch; and event filters narrow the feasible event/activity pairs.
Operational checks such as \texttt{success()} and predicates over unresolved runtime values such as step outputs are treated conservatively as \texttt{passable}.
Negation (logical \texttt{NOT} applied to an atomic predicate or guard subtree) is also handled conservatively because inverting an access-control predicate may enlarge, rather than shrink, the attacker's feasible set.
For reporting, the guard-filtered candidate set is summarized as
\[
\mathit{baseline}(n) = \textsc{BestMode}(C_n^{\mathsf{guard}}),
\]
where \textsc{BestMode} returns the highest-autonomy surviving mode under the reporting precedence above and returns \texttt{none} for the empty set.

\noindent\textbf{Runtime Feasibility Filters.}
Guard reachability is necessary but not sufficient: a node may still be inert if the trigger context lacks the secrets needed to activate the model call, or if the action enforces its own actor policy independently of workflow guards.
\tool therefore applies two additional filters:
\[
C_n^{\mathsf{act}} = \textsc{ActivationGate}(C_n^{\mathsf{guard}},\; \mathsf{secretReq}(n)),
\]
\[
C_n = \textsc{PolicyGate}(C_n^{\mathsf{act}},\; n,\; \mathcal{P}).
\]
The activation gate removes low-privilege trigger candidates when $\mathsf{secretReq}(n)=\mathsf{true}$ but the corresponding GitHub trigger context would not expose the needed secrets.
The policy gate models action-level access controls, such as \texttt{allowed\_non\_write\_users} in anthropics/claude-code-action~\cite{githubGitHubAnthropicsclaudecodeaction} and \texttt{allow-users} in openai/codex-action~\cite{githubGitHubOpenaicodexaction}.
If the invocation does not explicitly permit external users, \tool removes direct trigger candidates for that node.

\noindent\textbf{Target-Node Outcome.}
The final surviving trigger set is $C_n$.
For characterization, \tool summarizes it as
\[
\mathit{mode}(n) = \textsc{BestMode}(C_n),
\]
and the analysis returns
\[
\mathit{Triggerability}(n) = \langle C_n,\; \mathit{baseline}(n),\; \mathit{mode}(n) \rangle.
\]
If $C_n = \emptyset$, then $\mathit{mode}(n) = \texttt{none}$ and the per-step reachability map is empty.

\subsubsection{Per-Step Reachability Map}\label{sec:trigger:reach}

To support later dataflow reasoning, \tool next evaluates step reachability over the analysis region
\[
R(n) = \mathrm{ancestors}(J_n) \cup \{J_n\} \cup \mathrm{descendants}(J_n).
\]
Here, both \(\mathrm{ancestors}\) and \(\mathrm{descendants}\) denote transitive closure over job-level \texttt{needs} edges rather than one-hop neighbors.
For each step $s \in R(n)$, \tool reuses the same guard evaluator on the step's full prerequisite scope
\[
G(s) = \mathrm{ancestors}(J_s) \cup \{J_s\},
\]
together with $s$'s own step-level guard, yielding
\[
C_{\mathrm{raw}}(s) = \textsc{ReachableTriggers}(s,\; G(s)).
\]
Unlike target-node triggerability, this step-local pass includes $s$'s own guard because its purpose is to determine whether that specific step can execute under a given trigger context.
\tool then conditions the step on the surviving target-node context via
\[
C_{\mathrm{eff}}(s,n) = C_{\mathrm{raw}}(s) \cap C_n.
\]
The resulting map stores, for each step,
\[
\mathit{StepReachability}(n)[s] = \bigl(C_{\mathrm{raw}}(s),\; C_{\mathrm{eff}}(s,n)\bigr).
\]
$C_{\mathrm{raw}}(s)$ answers ``under which externally triggerable workflow events could this step run at all?'', while $C_{\mathrm{eff}}(s,n)$ answers ``under which of the events that actually reach this \ac{llm} node could this step also run?''.
Later stages restrict step summarization and taint propagation to steps with $C_{\mathrm{eff}}(s,n) \neq \emptyset$.

\subsection{Dataflow Analysis}\label{sec:dataflow}

We adopt a hybrid dataflow analysis design, which leverages \ac{llm}-assisted summarization for each relevant step locally, then propagates attacker and response taint deterministically across \ac{lwpg}.
For each triggerable \ac{llm} node $n$, our analysis answers four questions:
(1)~can attacker-controlled data reach the node's
prompt?
(2)~can the \ac{llm}'s output reach a security-relevant sink?
(3)~does attacker-controlled data also reach a direct execution sink co-present with the tainted prompt?
and (4)~does an attacker-controlled workspace feed the prompt?

\noindent\textbf{Analysis Region.}
For an \ac{llm} node $n$ in job~$J_n$, \tool defines the analysis region $R(n) = \mathrm{ancestors}(J_n) \cup \{J_n\} \cup \mathrm{descendants}(J_n)$, where both \(\mathrm{ancestors}\) and \(\mathrm{descendants}\) denote transitive closure over \texttt{needs} edges.
The resulting jobs are processed in topological order.
The upstream portion of $R(n)$ captures values that may carry attacker data into the \ac{llm} job from prerequisite jobs;
the downstream portion captures paths along which the \ac{llm} response may reach security-relevant sinks in later jobs.
Within each job, steps are processed in YAML-declared order.
For downstream steps, guard evaluation may additionally inspect prerequisite jobs outside~$R(n)$ when those jobs are required to decide whether that downstream job can run;
$R(n)$ remains the summary and propagation region, while guard reasoning uses the step's full prerequisite scope.
Only steps with $C_{\mathrm{eff}}(s,n) \neq \emptyset$ (drawn from the $\mathit{StepReachability}(n)$ map in \autoref{sec:trigger:reach}) are summarized and allowed to participate in taint propagation for node~$n$.

Operationally, the analysis first summarizes each reachable step in $R(n)$, then performs forward taint propagation over the same region to produce the structured evidence.
Pseudocode for this procedure appears in \autoref{alg:dataflow-analysis} in Appendix~\ref{app:dataflow-procedure}.
\autoref{sec:method:per-step-dataflow-summaries} describes step summarization, which assigns each executable step in $R(n)$ a symbolic list of inflows and outflows by combining deterministic extraction with \ac{llm}-assisted semantic analysis.
\autoref{sec:method:forward-taint-propagation} describes forward taint propagation, which propagates attacker taint~$\mathcal{T}$ and \ac{llm}-response taint~$\mathcal{L}$ through inter-step carriers across the \ac{lwpg} and returns the dataflow view consumed by risk synthesis.

\subsubsection{Step Summarization}\label{sec:method:per-step-dataflow-summaries}

Each executable step $s \in R(n)$ with $C_{\mathrm{eff}}(s,n) \neq \emptyset$ is assigned a \emph{step summary}: a list of typed \emph{inflows} and \emph{outflows} that capture its direct data dependencies and outputs symbolically.

Each inflow records a canonical source expression (e.g., \texttt{github.event.comment.body}), a source type, and a delivery method; tracked attacker-controlled source expressions are listed in \autoref{tab:taint-sources} (Appendix~\ref{app:dataflow-tables}).
We distinguish four delivery methods for attacker-controlled data:
\emph{M1}~(YAML splice), where GitHub interpolates \texttt{\$\{\{ expr \}\}} expressions into scripts or action inputs before execution;
\emph{M2}~(actions API), where actions read event data or inputs through GitHub Actions libraries~\cite{githubToolkitpackagesgithubMain,githubToolkitpackagescoreMain};
\emph{M3}~(process environment), where steps read runner environment variables such as \texttt{\$GITHUB\_EVENT\_PATH};
and \emph{M4}~(explicit API fetch), where workflow code fetches data explicitly through GitHub clients such as \texttt{gh}~\cite{githubGitHubgh} or Octokit~\cite{githubOctokit}.
Each outflow records a canonical sink expression, a sink type drawn from \autoref{tab:sink-taxonomy} (Appendix~\ref{app:dataflow-tables}), and the inflows that contribute to it.
The sink taxonomy distinguishes execution sinks, decision sinks, content sinks, and propagation sinks, together with the virtual \ac{llm} boundary sinks \texttt{llm\_prompt}, \texttt{llm\_response}, and \texttt{agent\_control}.
The target step~$s_n$ is summarized in \emph{llm-target} mode so that these virtual boundary outflows are made explicit; all other steps use \emph{ordinary} mode.

\noindent\textbf{Deterministic M1 Extraction.}
For \texttt{run} steps, \tool scans the raw script for every \texttt{\$\{\{\ expr\ \}\}} token and emits an inflow for the expression together with a \texttt{yaml\_splice\_exec} outflow.
The execution outflow is unconditional because GitHub interpolates the expression value directly into the shell command string before the runner executes it.
For \texttt{uses} steps, \texttt{\$\{\{\ expr\ \}\}} tokens in \texttt{with} values yield inflows.
but no automatic execution outflow, because the interpolated value flows into an action input rather than directly into a shell.
This deterministic pass relies on pattern matching and requires no \ac{llm} call.

\noindent\textbf{LLM-Assisted Step Summarization.}
For flows not captured by this deterministic pass, \tool relies on an \ac{llm}-assisted summarizer.
For \texttt{run} steps, the summarizer receives the normalized script body; \tool also resolves one hop of locally invoked scripts and includes their content as supporting evidence.
For \texttt{uses} steps, the summarizer receives the action reference, resolved \texttt{with} inputs, the \texttt{action.yml} manifest, and an excerpt of the action's source code.
As JavaScript action logic is encapsulated, \tool packs focused evidence around high-signal program points, including GitHub input/output APIs, event-payload reads, environment reads, workspace file I/O, explicit execution calls, \ac{llm} SDK boundaries, and GitHub write operations.
This evidence-packing strategy lets the summarizer recover effective dataflow logic without full static analysis of action dependencies.
The \ac{llm} prompt is provided in our supplementary repository.

\noindent\textbf{Workspace Contamination.}
When \tool encounters an \texttt{actions/checkout} step that checks out a PR head ref, it emits a \emph{workspace contamination scope} rather than an ordinary summary.
This scope marks the workspace as attacker-controlled from the immediately following step through the remainder of the job, so later workspace reads carry attacker taint; if such a read contributes to the target step's \texttt{llm\_prompt} outflow, \tool additionally sets~$\mathit{workspace\_to\_llm\_prompt}$.

\noindent\textbf{Target-Step Semantic Metadata.}
The target step~$s_n$ is summarized in \emph{llm-target} mode, which extends the ordinary contract with target-specific metadata.
Besides explicit prompt and response outflows, the summary records whether the node acts as a \emph{security-decision-maker}, \emph{sensitive-content-generator}, or \emph{agent}; these roles are derived deterministically from the emitted outflows and are consumed directly by risk synthesis.
Concretely, model-mediated outputs to security gates induce the \emph{security-decision-maker} role, outputs to persistent GitHub surfaces induce the \emph{sensitive-content-generator} role, and an exposed \texttt{agent\_control} outflow or non-empty capability profile induces the \emph{agent} role.
For autonomous agents, the same target summary may additionally preserve visible capability metadata, such as available tools and approval mode.

\subsubsection{Forward Taint Propagation}\label{sec:method:forward-taint-propagation}

With all summaries in place, \tool performs forward taint propagation over $R(n)$ in topological order.
Attacker taint $\mathcal{T}$ is seeded from inflows whose source expressions map to the canonical attacker-controlled fields in \autoref{tab:taint-sources}.
\ac{llm}-response taint $\mathcal{L}$ is seeded from the \texttt{llm\_response} outflow of~$s_n$ once prompt taint is confirmed, and then propagated only through explicit response-carrying or model-mediated outflows emitted by the target summary.
Both taint sets are propagated through five carrier types: step outputs, job outputs, environment variables, workspace content, and artifacts.
Crucially, each carrier is annotated with the subset of trigger candidates under which it exists.
When a later step consumes that carrier, \tool intersects the producer's carrier trigger set with $C_{\mathrm{eff}}(s,n)$ for the consumer step; taint only propagates when the same workflow trigger context can witness both the producer and the consumer.
Workspace contamination scopes are handled the same way: a contaminated workspace read only contributes attacker taint when the contamination scope and the current step are both reachable under a shared trigger subset.
After processing each job, \tool refreshes job outputs so that downstream job output references correctly inherit taint.

\noindent\textbf{Evidence Extraction.}
At step~$s_n$, \tool uses the explicit virtual boundary outflows to derive the structured evidence later consumed by risk synthesis.
If attacker taint reaches an outflow with \texttt{sink\_type = llm\_prompt}, \tool records in $P_n$ the attacker-controlled source expressions that reach the prompt; this explicit prompt evidence is the basis on which \ac{llm}-response taint is seeded. If a contributing inflow is a workspace read under an active contamination scope, \tool additionally sets~$\mathit{workspace\_to\_llm\_prompt}$.
Independently, whenever attacker taint reaches an execution sink anywhere in $R(n)$, \tool records the responsible attacker-controlled source expressions in~$X_n$.
This source-indexed execution evidence is kept distinct from response-derived execution evidence so that later synthesis can test source identity rather than only existential taint.
For response egress, any target-step outflow whose contributing inflows include \texttt{llm\_response}, as well as any outflow annotated \texttt{is\_model\_mediated\,=\,true}, is treated as response-derived; this lets \tool set whether \ac{llm}-response taint reaches execution, decision, and content sinks through either same-step or downstream flows.
For agent nodes, if the virtual \texttt{agent\_control} sink receives attacker taint, \tool sets $\mathit{agent\_control\_tainted}$ while preserving any visible agent capability metadata on the target node.

\noindent\textbf{Dataflow View.}
Rather than classifying threat vectors directly in this stage, \tool returns the structured view
\[
\mathit{DataflowView}(n) =
\left\langle
\begin{array}{l}
P_n,\; X_n,\; W_n,\; A_n, \\
R_n^{\mathsf{exec}},\; R_n^{\mathsf{dec}},\; R_n^{\mathsf{cont}},\; \mathit{Roles}(n)
\end{array}
\right\rangle,
\]
where $P_n$ and $X_n$ are the source sets described above, $W_n$ denotes \texttt{workspace\_to\_llm\_prompt}, $R_n^{\mathsf{exec}}$, $R_n^{\mathsf{dec}}$, and $R_n^{\mathsf{cont}}$ denote whether \ac{llm}-response taint reaches execution, decision, and content sinks, respectively, $A_n$ denotes \texttt{agent\_control\_tainted}, and $\mathit{Roles}(n)$ is the semantic role set emitted by the target-step summary.

\subsection{Risk Synthesis}\label{sec:synthesis}

Generally, risk synthesis is defined over the pair
\[
\langle \mathit{Triggerability}(n), \mathit{DataflowView}(n) \rangle
\]
for each analyzed \ac{llm} node~$n$.
We first define the following derived predicates:
\[
\begin{aligned}
\mathit{Triggerable}(n) &\equiv \mathit{Triggerability}(n).\mathit{mode} \neq \texttt{none}, \\
\mathit{PromptTainted}(n) &\equiv P_n \neq \emptyset, \\
\mathit{SharedPromptExec}(n) &\equiv P_n \cap X_n \neq \emptyset.
\end{aligned}
\]
Let $\mathit{JudgeRole}(n)$, $\mathit{ContentRole}(n)$, and $\mathit{AgentRole}(n)$ denote whether the target-step summary assigns, respectively, the \emph{security-decision-maker}, \emph{sensitive-content-generator}, and \emph{agent} roles to~$n$.

\noindent\textbf{Formal Threat-Vector Rules.}
Using these predicates, \tool defines TV1--TV6 as:
\[
\begin{aligned}
\mathit{TV}_1(n) &\equiv \mathit{Triggerable}(n) \land \mathit{JudgeRole}(n) \\
&\quad \land \mathit{PromptTainted}(n) \land R_n^{\mathsf{dec}}, \\
\mathit{TV}_2(n) &\equiv \mathit{Triggerable}(n) \land \mathit{ContentRole}(n) \\
&\quad \land \mathit{PromptTainted}(n) \land R_n^{\mathsf{cont}}, \\
\mathit{TV}_3(n) &\equiv \mathit{Triggerable}(n) \land \mathit{AgentRole}(n) \land A_n, \\
\mathit{TV}_4(n) &\equiv \mathit{Triggerable}(n) \land \mathit{SharedPromptExec}(n), \\
\mathit{TV}_5(n) &\equiv \mathit{Triggerable}(n) \land \mathit{PromptTainted}(n) \land R_n^{\mathsf{exec}}, \\
\mathit{TV}_6(n) &\equiv \mathit{Triggerable}(n) \land W_n.
\end{aligned}
\]
The key distinction between TV4 and TV5 is that TV4 requires source identity preservation across both the prompt path and a direct execution sink, while TV5 requires attacker-tainted prompt ingress together with response-derived execution.

\noindent\textbf{TV7 Heuristic.}
Unlike TV1--TV6, TV7 does not correspond to a demonstrated taint chain.
We therefore report it using the separate exposure heuristic
\[
\mathit{TV}_7(n) \equiv \mathit{Triggerability}(n).\mathit{mode} = \texttt{direct}.
\]
This heuristic flags \ac{llm} nodes with attacker-autonomous (\texttt{direct}) triggerability as susceptible to token-consumption abuse.
Nodes with \texttt{mediated} or \texttt{deferred} triggerability may incur incidental token consumption, but because the attacker cannot cause repeated invocations unilaterally, they do not warrant a TV7 finding.

%% file: sections/evaluation.tex
\section{Evaluation}\label{sec:eval}

In this section, we evaluate \tool by answering the following research questions (RQs):

\begin{itemize}
    \item \textbf{RQ1:} What is the landscape of \ac{llm}-integrated \ac{ci} workflows in real-world GitHub repositories? (\autoref{sec:eval-ecosystem})
    \item \textbf{RQ2:} How accurately does \tool detect \ac{llm}-induced security risks? (\autoref{sec:eval-detect})
    \item \textbf{RQ3:} What real-world findings does \tool uncover; what do concrete vulnerabilities reveal? (\autoref{sec:eval-findings})
\end{itemize}

\subsection{Implementation}\label{sec:impl}

We implement \tool in Python.
The symbolic backbone is deterministic, including workflow parsing, graph construction, guard evaluation, and taint propagation, while \acp{llm} are used only for semantic lifting tasks such as \ac{llm}-node identification, activation profiling, and step summarization.
We use gemini-3-flash-preview model~\cite{googleGeminiFlash} with temperature set to 0 and \texttt{reasoning\_effort} set to \texttt{medium}.

\subsection{Dataset Preparation}\label{sec:dataset}

We first build a broad GitHub workflow corpus following prior work~\cite{muralee2023argus,koishybayev2022characterizing}.
Specifically, we start from GH Archive event records~\cite{gharchiveArchive}, extract repositories whose activities beginning from Jan 2025 includes \texttt{github-actions[bot]}, and download the latest snapshot of each default branch.
From each repository, we collect valid YAML files under \texttt{.github/workflows/}, yielding about $5.2$ million workflow files corresponding to around $1.5$ million unique workflow content hashes.
We then derive an \ac{llm}-related candidate workflow set using two complementary high-recall signals.
First, we match a curated list of $53$ \ac{llm} API-key keywords
against workflow contents.
Second, to capture workflows that invoke \acp{llm} through dedicated actions, we curate a list of $467$ \ac{llm}-interaction actions from the GitHub Marketplace~\cite{githubGitHubMarketplace} and manual investigation, and select workflows that use such actions.
Merging the two result sets yields a candidate dataset of $61,606$ workflows associated with potential \ac{llm} usage, with $21,087$ unique content hashes, from $32,316$ repositories.

Because these retrieval signals are designed for broad coverage rather than
semantic confirmation, some collected workflows do not contain an actual \ac{llm}-interaction node.
We therefore run \tool's \ac{llm}-node identification over this candidate dataset and retain only workflows with at least one identified \ac{llm}-interaction node.
This filtering yields $50,354$ workflow instances with $16,818$ unique content hashes from $29,202$ repositories, which we treat as the \ac{llm}-integrated subset in the subsequent evaluations.
\autoref{table:dataset-filtering} summarizes how the dataset size changes across retrieval and filtering stages.

\input{tables/dataset-changes}

\subsection{Ecosystem Characterization}\label{sec:eval-ecosystem}

\input{tables/top-10-actions}

\input{tables/top-10-api-keys}

\input{tables/top-10-events}

This subsection characterizes the $16,818$ content-unique \ac{llm}-integrated workflow specifications.
Overall, this ecosystem is organized around recurring integration patterns rather than isolated one-off designs. 
Across the corpus, we observe $27,111$ jobs and $220,748$ steps, averaging $1.61$ jobs and $13.13$ steps per workflow specification.
\tool identifies $25,351$ \ac{llm} nodes in total, or $1.51$ per workflow on average.
While $66.3\%$ of workflows contain a single \ac{llm} node, the remaining $33.7\%$ contain multiple \ac{llm} nodes.

The ecosystem is also concentrated along three visible dimensions.
\autoref{table:action_stats} shows that a small set of assistants dominates action-based integrations, led by anthropics/claude-code-action, while some lower-rank actions exhibit strong template-style reuse.
\autoref{table:keyword_stats} shows a similar concentration in credentials, with OpenAI-, Anthropic-, and Gemini-related keys accounting for most references.
\autoref{table:workflow_stats} shows that externally influenced events such as \texttt{issues}, \texttt{pull\_request}, and \texttt{issue\_comment} remain highly prevalent, alongside indirect triggers such as \texttt{schedule} and \texttt{workflow\_call}.

\begin{figure}[t]
    \centering
    \includegraphics[width=\linewidth]{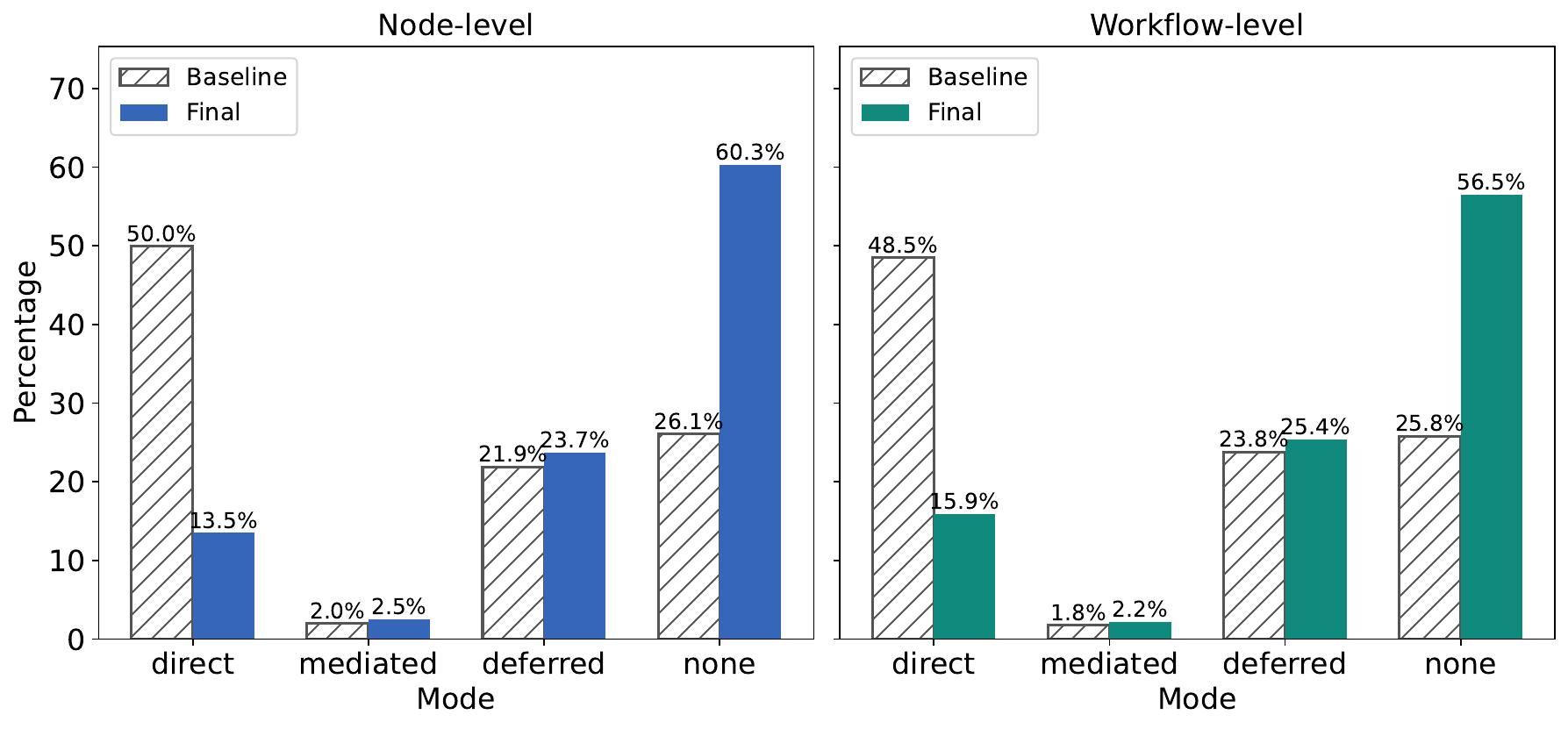}
    \caption{Baseline and final triggerability mode distributions at the node and workflow levels. Hatched bars show baseline modes before runtime feasibility filtering; solid bars show the final modes after triggerability analysis.}
    \label{fig:rq1-triggerability}
\end{figure}

As shown in \autoref{fig:rq1-triggerability}, \texttt{direct} is the dominant baseline mode at both the node level ($50.0\%$) and the workflow level ($48.5\%$), but many of these cases collapse after runtime feasibility filtering, most often to \texttt{none}.
\begin{findingbox}
\textbf{Finding.} $39.7\%$ of \ac{llm} nodes are externally triggerable, and $43.5\%$ of workflows contain at least one externally triggerable \ac{llm} node.
\end{findingbox}

\subsection{Detection Effectiveness}\label{sec:eval-detect}

To evaluate detection accuracy, we manually annotate a stratified sample of content-unique workflow specifications and compare \tool's outputs against the human-labeled ground truth.
This evaluation is designed to measure both component-level correctness and end-to-end threat-vector detection quality.
Unless otherwise noted, TV7 is excluded from standard precision/recall metrics because \tool reports it using a triggerability-based exposure heuristic rather than a demonstrated taint chain.

\subsubsection{Evaluation Setup}\label{sec:eval-detect-setup}

We sample $300$ content-unique workflow specifications using stratified sampling over \tool's predicted outputs.
The objective is to obtain sufficient coverage for both component-level error analysis and end-to-end threat-vector evaluation within a fixed annotation budget.
The benchmark allocates annotation effort across three predicted-outcome regions: workflows with no externally triggerable \ac{llm} node, workflows with at least one externally triggerable \ac{llm} node but no predicted TV1--TV6 finding, and workflows with at least one predicted TV1--TV6 finding.

We define three disjoint strata.
\textbf{S1} ($n = 50$): workflows for which \tool predicts no externally triggerable \ac{llm} node.
This stratum tests whether \tool's triggerability analysis produces false negatives, i.e., whether it incorrectly suppresses attacker reachability.
\textbf{S2} ($n = 50$): workflows for which \tool predicts at least one externally triggerable \ac{llm} node but no TV1--TV6 finding.
This stratum tests whether \tool misses threat vectors in workflows it correctly identifies as externally exposed.
\textbf{S3} ($n = 200$): workflows for which \tool predicts at least one TV1--TV6 finding.
This is the primary evaluation stratum for threat-vector detection quality; the larger quota reflects that all TV precision and recall signal resides here, and that coverage of rare threat vectors requires a larger sample.
Within each stratum, workflows are selected by uniform random sampling from the full corpus.
Three annotators with expertise in \ac{llm} security and GitHub \ac{ci} workflows independently labeled all $300$ workflows, with disagreements resolved by majority vote.

\subsubsection{LLM-Node Identification}\label{sec:eval-detect-llm}

We first evaluate whether \tool correctly identifies the workflow steps that
perform \ac{llm} interactions.
We compare the set of \ac{llm} nodes predicted by \tool against the ground-truth node sets by the human annotators, measuring step-level precision, recall, and F1 across all $300$ evaluation workflows.
\tool achieves near-perfect step-level identification (\autoref{table:llm-node-id}), with a recall of $100\%$ and precision of $98.8\%$.
The $5$ false positives (\textrm{FP}) across $300$ workflows share the same root cause: \tool includes the scaffolding steps automatically injected by \texttt{anthropics/claude-code-action@beta}.
The human annotators excluded these steps as infrastructure rather than \ac{llm}-interaction steps.
There are zero false negatives (\textrm{FN}) on this task.

\begin{table}[t]
\centering
\caption{\ac{llm}-node identification accuracy at the step level.}
\label{table:llm-node-id}
\begin{tabular}{lrrrrrr}
\toprule
\textbf{Stratum} & \textbf{TP} & \textbf{FP} & \textbf{FN} & \textbf{Prec.} & \textbf{Rec.} & \textbf{F1} \\
\midrule
S1 ($n=50$) & 66 & 2 & 0 & 0.971 & 1.000 & 0.985 \\
S2 ($n=50$) & 75 & 0 & 0 & 1.000 & 1.000 & 1.000 \\
S3 ($n=200$) & 282 & 3 & 0 & 0.990 & 1.000 & 0.995 \\
\midrule
\textbf{Overall} & \textbf{423} & \textbf{5} & \textbf{0} & \textbf{0.988} & \textbf{1.000} & \textbf{0.994} \\
\bottomrule
\end{tabular}
\end{table}

\subsubsection{Triggerability Analysis}\label{sec:eval-detect-trig}

We evaluate whether \tool correctly determines how an external attacker can cause an \ac{llm} node to execute.
We compare the triggerability modes assigned by \tool against the ground-truth modes on all $423$ matched nodes (those identified as \ac{llm}-interaction steps by both \tool and the annotators).
\tool achieves $99.8\%$ exact mode accuracy, with only a single misclassification across all $423$ nodes.
\autoref{table:trig-confusion} shows the confusion matrix; the sole misclassification is one node where the human labeled the mode as \texttt{none} but \tool assigned \texttt{direct}.
This is a permissive (over-approximating) error caused by a custom step that performs write-permission checking and passes the result to a downstream \texttt{if:} condition, which is a guard pattern \tool does not model, so it conservatively assumes the condition does not block execution.
There are zero conservative errors: \tool never suppresses attacker reachability where the human annotator finds a triggerable node.
At the binary triggerability level (externally triggerable vs.\ not), \tool achieves $100\%$ recall, $99.7\%$ precision, and $99.8\%$ F1.

\begin{table}[t]
\centering
\caption{Triggerability mode confusion matrix (rows = human, cols = \tool).}
\label{table:trig-confusion}
\begin{tabular}{lrrrr}
\toprule
 & \texttt{none} & \texttt{direct} & \texttt{mediated} & \texttt{deferred} \\
\midrule
\texttt{none}     & 94 &   1 &  0 & 0 \\
\texttt{direct}   &  0 & 287 &  0 & 0 \\
\texttt{mediated} &  0 &   0 & 39 & 0 \\
\texttt{deferred} &  0 &   0 &  0 & 2 \\
\bottomrule
\end{tabular}
\end{table}

\subsubsection{Threat-Vector Detection}\label{sec:eval-detect-tv}

We then evaluate the end-to-end detection of TV1--TV6 by comparing \tool's synthesized threat-vector findings against the ground-truth labels.
This stage captures the combined effect of node identification, triggerability reasoning, and dataflow-aware risk synthesis.
Threat-vector detection is evaluated as a multi-label workflow-level classification task over TV1--TV6, since a single workflow may exhibit multiple threat vectors simultaneously.
For each TV, we score a binary one-vs-rest task: a workflow is predicted positive if \tool reports that TV on at least one node, and the ground truth is positive if any human-annotated node carries that TV.

\autoref{table:tv-detection} reports the resulting per-TV workflow-level precision, recall, and F1, together with the positive/negative label support on our stratified 300-workflow benchmark.
The micro-average aggregates TP/FP/FN over all six TV labels and therefore emphasizes overall label-instance performance, whereas the macro-average gives equal weight to each TV and thus better reflects balance across common and rare threat vectors.
Because the benchmark is stratified by predicted outcome, these supports are evaluation supports rather than estimates of ecosystem prevalence.

\begin{findingbox}
\textbf{Finding.} Overall, \tool achieves a micro-average F1 of $0.917$ (P\,=\,$0.934$, R\,=\,$0.900$) and a macro-average F1 of $0.874$ on TV1--TV6.
\end{findingbox}

The main source of reduced overall recall is the S2 blind spot: by construction, S2 workflows are those for which \tool predicts an externally triggerable \ac{llm} node but no TV1--TV6 finding, yet human annotators identify $25$ of the $50$ S2 workflows as carrying at least one TV1--TV6 risk.
These cases drive the majority of FNs across all TVs, particularly for TV1 ($15$ FNs) and TV4 ($10$ FNs).
On the precision side, TV5 is the primary source of over-prediction ($20$ FPs).
TV1, TV2, and TV6 achieve perfect or near-perfect precision.

\begin{table}[t]
\centering
\caption{Per-TV workflow-level detection accuracy (TV1--TV6). Pos./Neg.\ denote ground-truth positive/negative workflow counts on the stratified 300-workflow benchmark.}
\label{table:tv-detection}
\begin{tabular}{lrrrrrrrr}
\toprule
\textbf{TV} & \textbf{Pos.} & \textbf{Neg.} & \textbf{TP} & \textbf{FP} & \textbf{FN} & \textbf{Prec.} & \textbf{Rec.} & \textbf{F1} \\
\midrule
TV1 &  33 & 267 &  18 &  0 & 15 & 1.000 & 0.545 & 0.706 \\
TV2 & 145 & 155 & 137 &  3 &  8 & 0.979 & 0.945 & 0.961 \\
TV3 &  62 & 238 &  53 &  2 &  9 & 0.964 & 0.855 & 0.906 \\
TV4 & 110 & 190 & 100 &  6 & 10 & 0.943 & 0.909 & 0.926 \\
TV5 & 122 & 178 & 120 & 20 &  2 & 0.857 & 0.984 & 0.916 \\
TV6 &  17 & 283 &  12 &  0 &  5 & 1.000 & 0.706 & 0.828 \\
\midrule
\textbf{Micro} & 489 & 1311 & 440 & 31 & 49 & 0.934 & 0.900 & 0.917 \\
\textbf{Macro} & --- & --- & --- & --- & --- & 0.957 & 0.824 & 0.874 \\
\bottomrule
\end{tabular}
\end{table}

\subsection{End-to-End Findings}\label{sec:eval-findings}

To assess real-world findings, we apply \tool to the \ac{llm}-integrated workflow corpus and conduct a responsible disclosure campaign targeting TV4--TV6 findings.
These three threat vectors correspond to concrete, attacker-feasible exploit paths with clear repository-side remediations, making them directly actionable.
TV1--TV3 and TV7 capture higher-level behavioral or systemic conditions whose mitigations are more context-dependent; we report them as part of the broader risk landscape but do not include them in the disclosure campaign.
Detection and disclosure are ongoing: as new workflows are analyzed, confirmed findings continue to be reported to affected repository maintainers.

\subsubsection{Threat-Vector Prevalence}

\begin{figure}[t]
    \centering
    \includegraphics[width=0.95\linewidth]{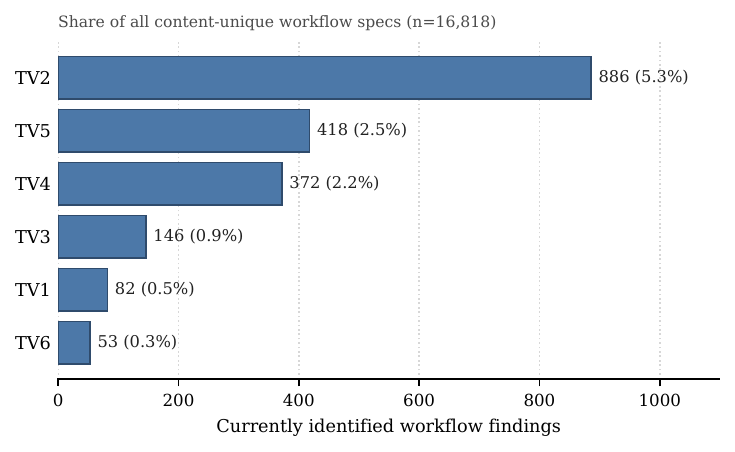}
    \caption{Lower-bound prevalence of currently identified TV1--TV6 findings over all $16{,}818$ content-unique workflow specifications in the ongoing end-to-end scan.}
    \label{fig:rq3-tv-distribution}
\end{figure}

Because end-to-end scanning is still ongoing, we report current lower-bound prevalence over the full set of $16,818$ content-unique workflow specifications rather than finalized corpus-wide rates.
As shown in \autoref{fig:rq3-tv-distribution}, TV2 currently appears in $5.3\%$ of workflows, followed by TV5 ($2.5\%$) and TV4 ($2.2\%$), while TV3 ($0.9\%$), TV1 ($0.5\%$), and TV6 ($0.3\%$) are less common.
We omit TV7 from the figure because it is a triggerability-based heuristic rather than a demonstrated taint-chain finding; nonetheless, the current scan has already identified $3,731$ workflow specifications ($22.2\%$) with TV7 exposure.

\subsubsection{Responsible Disclosure}
Before disclosure, we manually verify each detected TV4--TV6 case to confirm it exposes a real vulnerability, and disclose only this verified subset to minimize unnecessary burden on repository maintainers.
\begin{findingbox}
\textbf{Finding.} As of April 30, 2026, we have disclosed $802$ workflow instances ($351$ content-unique workflow specifications) across $759$ repositories, of which \acknum have acknowledged the report and fixed the issues.
\end{findingbox}
Acknowledgment counts should be interpreted as lower bounds, as maintainer response times vary and the disclosure process is ongoing.
Among the disclosed repositories, $6$ have over 1K GitHub stars, including $2$ with over 10K stars, indicating that the identified vulnerabilities affect widely-used, actively maintained projects.

\subsubsection{Case Study}\label{sec:case-study}

\input{code/case-study}

We illustrate \tool's end-to-end analysis with a representative workflow that implements a Gemini \texttt{@mention} bot: any user who mentions \texttt{@gemini} in a GitHub issue comment triggers the workflow, which forwards the request to the Gemini \ac{llm}, applies the generated code changes to the workspace via an \texttt{awk} script, and can auto-merge the resulting pull request.
The key logic of this workflow is presented in \autoref{lst:case-study}.
The workflow holds a live \texttt{GEMINI\_API\_KEY} secret and grants \texttt{contents}, \texttt{pull-requests}, \texttt{issues}, and \texttt{actions} write permissions.
Because it triggers on \texttt{issue\_comment} with no author restriction, \tool assigns the \ac{llm} node a \texttt{direct} triggerability mode.

\tool detects five threat vectors on this workflow.
\textbf{TV2:} The \ac{llm}'s output is written to the workspace (line 23) and committed as a pull request; including \texttt{--auto-merge} in the issue text causes the PR to merge automatically (line 29), completing a fully attacker-controlled repository poisoning chain.
\textbf{TV3:} Issue comment body flows into the \ac{llm} prompt (line 22) and \texttt{agent\_control} sink (line 23); with the Gemini CLI running in \texttt{--yolo} mode, a crafted issue body can hijack the agent to perform arbitrary operations.
\textbf{TV4:} The same untrusted content is also interpolated via GitHub expressions directly into \texttt{run:} shell blocks without sanitization (line 10, 22), creating a direct execution-injection path before the \ac{llm} is even invoked.
\textbf{TV5:} The \ac{llm}'s response is parsed by an \texttt{awk} script that calls \texttt{system("mkdir ...")} on model-chosen paths (line 26), making an \ac{llm}-controlled value the argument to a shell execution call.
\textbf{TV7:} The \ac{llm} node is directly triggerable; if the repository owner has not configured external rate controls, an attacker can repeatedly open issues or post comments to exhaust the API quota.

%% file: tables/dataset-changes.tex
\begin{table}[t]
\centering
\small
\caption{Dataset size across retrieval and filtering stages. The first two rows are parallel high-recall retrieval signals; the merged candidate set is their union.}
\label{table:dataset-filtering}
\resizebox{0.9\linewidth}{!}{%
\begin{tabular}{lrrr}
\toprule
\textbf{Stage} & \textbf{Workflows} & \textbf{Repos.} & \textbf{Unique hashes} \\
\midrule
Keyword match & $54,201$ & $26,793$ & $20,255$ \\
Action match & $27,775$ & $17,191$ & $6,962$ \\
Merged candidates & $61,606$ & $32,316$ & $21,087$ \\
\ac{llm}-node filtering & $50,354$ & $29,202$ & $16,818$ \\
\bottomrule
\end{tabular}}
\end{table}

%% file: tables/top-10-actions.tex
\begin{table}[t]
\centering
\caption{Top-10 LLM-interaction actions in the dataset.}
\label{table:action_stats}
\small
\renewcommand{\arraystretch}{1.1}
\resizebox{\linewidth}{!}{%
\begin{tabular}{lrr}
\toprule
\textbf{Action} & \textbf{Total Frequency} & \textbf{Unique Frequency} \\ \midrule
anthropics/claude-code-action & 14,028 & 3,561 \\
johnsonran/nomore-spam & 3,867 & 8 \\
google-github-actions/run-gemini-cli & 2,719 & 668 \\
actions/ai-inference & 1,135 & 294 \\
\{qodo-ai,codium-ai\}/pr-agent & 771 & 371 \\
anthropics/claude-code-base-action & 364 & 153 \\
ultralytics/actions & 357 & 30 \\
anc95/chatgpt-codereview & 286 & 181 \\
openai/codex-action & 262 & 122 \\
0xjord4n/aixion & 207 & 7 \\
\bottomrule
\end{tabular}}
\end{table}

%% file: tables/top-10-api-keys.tex
\begin{table}[t]
\centering
\caption{Top-10 LLM API key keywords in the dataset.
  \colorbox{green!20}{Shaded green}: OpenAI keys.
  \colorbox{orange!25}{Shaded orange}: Anthropic keys.
  Unshaded: other providers.}
\label{table:keyword_stats}
\small
\renewcommand{\arraystretch}{1.1}
\resizebox{\linewidth}{!}{%
\begin{tabular}{lrr}
\toprule
\textbf{Keyword} & \textbf{Total Frequency} & \textbf{Unique Frequency} \\ \midrule
\rowcolor{green!20}\texttt{OPENAI\_API\_KEY} & 16,250 & 6,092 \\
\rowcolor{orange!25}\texttt{ANTHROPIC\_API\_KEY} & 11,203 & 3,745 \\
\rowcolor{orange!25}\texttt{CLAUDE\_CODE\_OAUTH\_TOKEN} & 8,589 & 2,196 \\
\texttt{GEMINI\_API\_KEY} & 6,625 & 3,120 \\
\texttt{OPENROUTER\_API\_KEY} & 1,527 & 716 \\
\rowcolor{green!20}\texttt{OPENAI\_KEY} & 1,077 & 435 \\
\texttt{GROQ\_API\_KEY} & 929 & 528 \\
\rowcolor{green!20}\texttt{AZURE\_OPENAI\_API\_KEY} & 860 & 335 \\
\texttt{DEEPSEEK\_API\_KEY} & 711 & 419 \\
\texttt{LLM\_API\_KEY} & 652 & 336 \\
\bottomrule
\end{tabular}}
\end{table}

%% file: tables/top-10-events.tex
\begin{table}[t]
\centering
\caption{Top-10 events triggering workflows in the dataset.
  \colorbox{red!15}{Shaded red}: triggerable by external (untrusted) users.
  \colorbox{blue!12}{Shaded blue}: triggered indirectly (scheduled or via workflow call).
  Unshaded: maintainer-only triggers.}
\label{table:workflow_stats}
\small
\renewcommand{\arraystretch}{1.1}
\resizebox{\linewidth}{!}{%
\begin{tabular}{lrr}
\toprule
\textbf{Event} & \textbf{Total Frequency} & \textbf{Unique Frequency} \\ \midrule
\texttt{workflow\_dispatch} & 20,908 & 8,183 \\
\rowcolor{red!15}\texttt{issues} & 16,609 & 2,915 \\
\rowcolor{red!15}\texttt{pull\_request} & 11,747 & 5,282 \\
\rowcolor{red!15}\texttt{issue\_comment} & 11,282 & 3,267 \\
\rowcolor{blue!12}\texttt{schedule} & 9,850 & 4,177 \\
\rowcolor{red!15}\texttt{pull\_request\_review\_comment} & 8,646 & 2,069 \\
\rowcolor{red!15}\texttt{pull\_request\_review} & 7,840 & 1,624 \\
\texttt{push} & 7,107 & 2,709 \\
\rowcolor{red!15}\texttt{pull\_request\_target} & 4,905 & 399 \\
\rowcolor{blue!12}\texttt{workflow\_call} & 3,954 & 771 \\
\bottomrule
\end{tabular}}
\end{table}

%% file: code/case-study.tex
\begin{lstlisting}[
  style=lst-top,
  captionpos=t,
  frame=single,
  framerule=0.4pt,
  aboveskip=0.5em,
  columns=flexible,
  showstringspaces=false,
  basicstyle={\scriptsize\ttfamily},
  numbers=left,
  numbersep=6pt,
  xleftmargin=2em,
  xrightmargin=1em,
  framexleftmargin=0pt,
  framexrightmargin=0pt,
  framesep=3pt,
  caption={Gemini \texttt{@mention} workflow (excerpt).},
  label={lst:case-study},
  numberstyle=\tiny\color{gray},
  breaklines=true,
  breakatwhitespace=true,
  linebackgroundcolor={%
    \ifnum\value{lstnumber}=10\color{blue!15}\fi%
    \ifnum\value{lstnumber}=22\color{blue!15}\fi%
    \ifnum\value{lstnumber}=23\color{blue!15}\fi%
    \ifnum\value{lstnumber}=26\color{blue!15}\fi%
    \ifnum\value{lstnumber}=29\color{blue!15}\fi%
  }
]
on:
  issue_comment: ...
permissions: ...
jobs:
  gemini_mention:
    if: contains(github.event.comment.body, '@gemini')
    steps:
      - name: Extract command
        run: |
          BODY="${{ github.event.comment.body }}"
          COMMAND=$(echo "$BODY" | grep -oP gemini ...)
          if echo "$COMMAND" | grep "auto-merge"; then
            echo "auto_merge=true" >> $GITHUB_OUTPUT
          fi
          echo "command=$COMMAND" >> $GITHUB_OUTPUT
      ...
      - name: Run Gemini
        env:
          GEMINI_API_KEY: ${{ secrets.GEMINI_API_KEY }}
        run: |
          PROMPT="You are a software engineer ...
            Req: ${{ steps.extract.outputs.command }}"
          gemini --yolo --prompt "$PROMPT" > g.md
      - name: Apply changes
        run: |
          awk '...system("mkdir \"" dir "\"")...' g.md
      - name: Enable auto-merge
        if: steps.extract.outputs.auto_merge == 'true'
        run: gh pr merge ...
\end{lstlisting}

%% file: sections/discussion.tex
\section{Discussion}\label{sec:discuss}

\subsection{Defensive Takeaways}\label{sec:discuss_defense}

\noindent\textbf{Principle of Least Privilege.}
Permissions should be declared at the job level and scoped to the minimum required (e.g., \texttt{issues: write} for a labeling job).
Besides, it is necessary to gate \ac{llm}-integrated jobs on the triggering actor's identity or role.
Beyond \ac{ci}-specific controls, established prompt injection mitigations are applicable here; 
we refer practitioners to other research~\cite{chen2025struq,hines2024defending,zhong2026attention,chen2025defending} for a comprehensive treatment.

\noindent\textbf{Safe Handling of \ac{llm} Inputs and Outputs.}
On the input side, untrusted event payloads should never be spliced into execution sites via GitHub expression syntax.
Instead, untrusted values should be passed through environment variables, which the shell treats as data rather than code.
On the output side, \ac{llm}-generated content must not be forwarded unchecked to sensitive sinks.
Appropriate sanitization or structural constraints on the expected response format should be enforced before the output is consumed downstream.

\noindent\textbf{Fail-Safe API Integration.}
Workflows must handle \ac{llm} API errors explicitly.
Providers may return non-2xx status codes to signal blocked or flagged requests (e.g., Azure OpenAI returns HTTP~400 on content-filter violations~\cite{microsoftContentFiltering}).
Workflows should fail closed in such cases, act on any harm or injection signals exposed in the response, and enforce \ac{llm} invocation quotas to bound exposure to denial-of-wallet attacks.

\subsection{Limitations}\label{sec:discuss_limit}

\noindent\textbf{Boundary of the Analyzable Closure.}
\tool reasons over the workflow YAML and its immediate analyzable closure.
It does not recursively chase transitive dependencies or fully unpack opaque execution boundaries such as dynamically fetched payloads.
This can hide \ac{llm} calls or downstream sinks and induce false negatives.

\noindent\textbf{Conservative Treatment of Dynamic Guards.}
Triggerability analysis explicitly models event declarations, actor-based checks, and many workflow guards, but conditions that depend on step or job outputs are treated conservatively.
This can over-approximate attacker reachability when admission logic is computed dynamically at runtime.

\noindent\textbf{LLM-assisted Dataflow Summarization.}
\tool relies on an \ac{llm}-assisted step summarizer to recover inflows, outflows, and \ac{llm}-interaction boundaries.
In these scenarios, precise end-to-end static analysis does not scale.
Summarization may miss subtle flows or sanitization logic.

%% file: sections/related.tex
\section{Related Work}\label{sec:related}

\subsection{Security Analysis of \ac{ci} Workflows}\label{sec:related-ci-sec}

Prior work studies security risks in \ac{ci} artifacts such as workflow configurations, scripts, and reusable actions.
Existing work reports pervasive misconfiguration and overprivileged \texttt{GITHUB\_TOKEN} use~\cite{gallaba2018use,vassallo2020configuration,koishybayev2022characterizing}, and demonstrates attacks including token leakage, privilege escalation, artifact backdooring, cryptomining, plugin hijacking, and cache poisoning~\cite{gu2023continuous,pan2023ambush,li2022robbery,li2024toward,gu2024more}.
Automated analyses focus on taint-style detection of unsafe data flows and least-privilege reduction~\cite{muralee2023argus,tystahl2026cosseter}.
In contrast, our threat model centers on \acp{llm} embedded in running workflows, which can be manipulated through attacker-controlled natural language.

\subsection{Prompt Injection in \ac{llm}-Integrated Applications}\label{sec:related-llm-sec}

Prompt injection in \ac{llm}-integrated applications has been studied extensively, from early demonstrations and attack taxonomies~\cite{greshake2023not,liu2024formalizing,kimsok2026sok} to concrete exploits against \ac{llm} frameworks and coding agents~\cite{liu2024demystifying,liu2025your}. 
Subsequent work expands the threat surface to tool selection and \ac{llm}-as-a-judge pipelines~\cite{shi2026prompt,shi2024optimization}, while detection and defense efforts use fuzzing, taint tracking, and prompt sanitization~\cite{liu2025make,he2026taintp2x,zhong2026attention}. 
However, these studies do not address \ac{ci} workflows, where privileged automation consumes untrusted inputs under a distinct operational and security model.

%% file: sections/conclusion.tex
\section{Conclusion}\label{sec:conclusion}

In this paper, we present the first study of \ac{llm}-induced security risks in GitHub \ac{ci} workflows, introducing a taxonomy of risk classes and threat vectors.
We then propose \tool, a hybrid analysis framework that achieves high accuracy in controlled evaluation and reveals that $43.5\%$ of real-world \ac{llm}-integrated workflows are externally triggerable.
Our responsible disclosure of $802$ vulnerable workflow instances across $759$ repositories shows that these threat vectors arise in real repositories:
they affect actively maintained projects and can be verified and remediated in practice.

%% file: sections/appendix.tex
\section{Scope and Assumptions}\label{app:scope}

Our analysis targets an \ac{llm}-integrated workflow together with its immediate \textit{analyzable closure}: the workflow YAML itself, repository-local scripts directly invoked by runners, directly referenced third-party \ac{ci} actions, and reusable workflows.
Several elements remain out of scope.
We do not model the proprietary internals or provider-specific safety semantics of remote \acp{llm}; perform unbounded recursive analysis of arbitrary transitive dependencies in third-party action ecosystems; or treat dynamically loaded execution elements, such as opaque Docker actions (\texttt{uses: docker://}) and network-fetched payloads (e.g., \texttt{curl | bash}), as explicit analyzable paths.
We also study repositories under GitHub's default security posture rather than configurations that intentionally disable core isolation guarantees, such as non-default setups that expose secrets to pull requests from arbitrary forks.

\section{Impact Taxonomy}\label{app:impact-taxonomy}

While the main text centers on risk classes and threat vectors, we retain an impact taxonomy as a supporting view that summarizes the harms associated with each threat vector.

\subsection{Security Impacts}

We define six categories of security impact.
\textbf{I1} (\textit{Execution Environment Compromise}) denotes unauthorized control over the \ac{ci} execution environment.
\textbf{I2} (\textit{Secret Exfiltration}) covers the unauthorized disclosure of sensitive credentials, such as API keys or GitHub tokens, to the adversary.
\textbf{I3} (\textit{Repository \& Workflow Tampering}) refers to attacker-driven modifications of repository contents, workflow definitions, or other persistent project state.
\textbf{I4} (\textit{Security Gate Bypass}) arises when security checks, content moderation, or reviews incorrectly approve attacker-controlled submissions.
\textbf{I5} (\textit{Misinformation Propagation}) captures cases where the workflow disseminates attacker-influenced contents.
Finally, \textbf{I6} (\textit{Availability Degradation}) occurs when adversary-induced \ac{llm} invocations exhaust token budgets or rate limits.

\subsection{Threat-Vector-to-Impact Mapping}\label{app:taxonomy:matrix}

The relationships among risk classes, threat vectors, and impacts are shown in \autoref{tab:taxonomy-matrix}.
A single threat vector may induce multiple impacts depending on the privilege level and workflow configuration of the concrete deployment.

\input{tables/taxonomy-mapping}

\section{Methodology Details}\label{app:method-details}

\subsection{Event/Activity Matrix}\label{app:event-matrix}

\autoref{tab:event-matrix} lists every GitHub event and activity type handled by \tool's baseline trigger classification, grouped by the assigned \texttt{TriggerMode}.
Events absent from this table receive \texttt{none} by default.

\begin{table}[t]
\centering
\caption{Complete event/activity matrix used by \tool for baseline trigger classification. Activity type \texttt{*} denotes the catch-all entry: any activity not explicitly listed for that event falls back to this row.}
\label{tab:event-matrix}
\footnotesize
\renewcommand{\arraystretch}{1.15}
\setlength{\tabcolsep}{4pt}
\resizebox{\linewidth}{!}{%
\begin{tabular}{llr}
\toprule
\textbf{Event} & \textbf{Activity Types} & \textbf{Mode} \\
\midrule
\multicolumn{3}{l}{\textit{Direct\,---\,attacker controls content immediately}} \\[1pt]
\texttt{issues}
  & \texttt{opened}, \texttt{edited}, \texttt{reopened}, \texttt{*}
  & \texttt{direct} \\
\texttt{issue\_comment}
  & \texttt{created}, \texttt{edited}, \texttt{*}
  & \texttt{direct} \\
\texttt{pull\_request}
  & \texttt{opened}, \texttt{edited}, \texttt{synchronize}, \texttt{reopened}, \texttt{*}
  & \texttt{direct} \\
\texttt{pull\_request\_target}
  & \texttt{opened}, \texttt{edited}, \texttt{synchronize}, \texttt{*}
  & \texttt{direct} \\
\texttt{pull\_request\_review\_comment}
  & \texttt{created}, \texttt{*}
  & \texttt{direct} \\
\texttt{pull\_request\_review}
  & \texttt{submitted}, \texttt{*}
  & \texttt{direct} \\
\texttt{discussion}
  & \texttt{created}, \texttt{edited}, \texttt{*}
  & \texttt{direct} \\
\texttt{discussion\_comment}
  & \texttt{created}, \texttt{*}
  & \texttt{direct} \\
\midrule
\multicolumn{3}{l}{\textit{Mediated\,---\,requires maintainer or reviewer action}} \\[1pt]
\texttt{issues}
  & \texttt{labeled}, \texttt{unlabeled}
  & \texttt{mediated} \\
\texttt{pull\_request}
  & \texttt{labeled}, \texttt{unlabeled}
  & \texttt{mediated} \\
\texttt{pull\_request\_target}
  & \texttt{labeled}, \texttt{unlabeled}
  & \texttt{mediated} \\
\texttt{pull\_request\_review}
  & \texttt{dismissed}
  & \texttt{mediated} \\
\midrule
\multicolumn{3}{l}{\textit{Deferred\,---\,attacker plants content; automated job triggers later}} \\[1pt]
\texttt{schedule}
  & \texttt{*}
  & \texttt{deferred} \\
\texttt{workflow\_run}
  & \texttt{*}
  & \texttt{deferred} \\
\midrule
\multicolumn{3}{l}{\textit{None\,---\,not externally triggerable}} \\[1pt]
\texttt{push}               & \texttt{*} & \texttt{none} \\
\texttt{workflow\_dispatch} & \texttt{*} & \texttt{none} \\
\texttt{repository\_dispatch} & \texttt{*} & \texttt{none} \\
\texttt{deployment}         & \texttt{*} & \texttt{none} \\
\texttt{deployment\_status} & \texttt{*} & \texttt{none} \\
\texttt{release}            & \texttt{*} & \texttt{none} \\
\bottomrule
\end{tabular}%
}
\end{table}

\subsection{Triggerability Procedure}\label{app:triggerability-procedure}

\autoref{alg:triggerability} presents compact pseudocode for the triggerability analysis described in \autoref{sec:trigger}.
The algorithm takes an \ac{llm} node~$n$ and the workflow's trigger declarations as input and returns $\mathit{Triggerability}(n) = \langle C_n,\, \mathit{baseline}(n),\, \mathit{mode}(n) \rangle$.

\input{tables/atomic-predicate}

\input{algorithms/triggerability-analysis}

\subsection{Guard Predicate Categories}\label{app:triggerability-details}

\autoref{tab:atomic-predicate} summarizes the atomic predicate categories used by \textsc{FilterByGuards} and \textsc{ReachableTriggers}.
The four categories partition the space of atomic boolean predicates that appear in GitHub workflow \texttt{if} conditions.

\subsection{Dataflow Procedure}\label{app:dataflow-procedure}

\autoref{alg:dataflow-analysis} presents compact pseudocode for the dataflow analysis described in \autoref{sec:dataflow}.
The algorithm operates over the analysis region $R(n)$ for a given \ac{llm} node~$n$, processing jobs in topological order and steps within each job in YAML-declared order.

\input{tables/taint-sources}

\input{algorithms/dataflow-analysis}

\input{tables/sink-taxonomy}

\subsection{Dataflow Reference Tables}\label{app:dataflow-tables}

\autoref{tab:taint-sources} lists the canonical attacker-controlled taint sources tracked by \tool, corresponding to the four delivery methods M1--M4 defined in \autoref{sec:method:per-step-dataflow-summaries}.
Any step inflow whose source expression matches a listed field is seeded with attacker taint~$\mathcal{T}$ during forward propagation.
\autoref{tab:sink-taxonomy} presents the complete sink taxonomy used by the step summarizer to type outflows.
The virtual boundary sinks \texttt{llm\_prompt}, \texttt{llm\_response}, and \texttt{agent\_control} are produced exclusively in \emph{llm-target} mode and serve as the evidence anchors for risk synthesis.

%% file: tables/taxonomy-mapping.tex
\begin{table}[t]
\caption{Unified taxonomy mapping. Each row represents a threat vector (TV1--TV7).
Colored units (\cunit) indicate the risk class(es) related to it (left) and the security impacts it may induce (right);
grey units (\eunit) indicate no direct relationship.}
\label{tab:taxonomy-matrix}
\centering
\renewcommand{\arraystretch}{1.2}
\resizebox{\linewidth}{!}{%
\begin{tabular}{ccccccc|c|cccccc}
\toprule
\multicolumn{7}{c|}{\textbf{Risk Class}}
  & \multirow{2}{*}{\textbf{TV}}
  & \multicolumn{6}{c}{\textbf{Impact}} \\
\cmidrule(r){1-7}\cmidrule(l){9-14}
R1 & R2 & R3 & R4 & R5 & R6 & R7 & & I1 & I2 & I3 & I4 & I5 & I6 \\
\midrule
\cunit & \eunit & \eunit & \eunit & \eunit & \eunit & \eunit & TV1 & \eunit & \eunit & \cunit & \cunit & \cunit & \eunit \\
\eunit & \cunit & \eunit & \eunit & \cunit & \eunit & \eunit & TV2 & \eunit & \eunit & \cunit & \eunit & \cunit & \eunit \\
\eunit & \eunit & \cunit & \cunit & \eunit & \eunit & \eunit & TV3 & \cunit & \cunit & \cunit & \eunit & \cunit & \eunit \\
\eunit & \eunit & \eunit & \cunit & \cunit & \eunit & \eunit & TV4 & \cunit & \cunit & \cunit & \eunit & \cunit & \eunit \\
\eunit & \eunit & \eunit & \cunit & \cunit & \eunit & \eunit & TV5 & \cunit & \cunit & \cunit & \eunit & \cunit & \eunit \\
\eunit & \eunit & \eunit & \cunit & \eunit & \cunit & \eunit & TV6 & \cunit & \cunit & \cunit & \eunit & \cunit & \eunit \\
\eunit & \eunit & \eunit & \eunit & \eunit & \eunit & \cunit & TV7 & \eunit & \eunit & \eunit & \eunit & \eunit & \cunit \\
\bottomrule
\end{tabular}}
\end{table}

%% file: tables/atomic-predicate.tex
\begin{table*}[t]
\centering
\caption{Atomic predicate categories used in triggerability analysis.}
\label{tab:atomic-predicate}
\small
\resizebox{\textwidth}{!}{
\begin{tabular}{lll}
\toprule
\textbf{Category} & \textbf{Security Interpretation} & \textbf{Representative Examples} \\
\midrule
\texttt{blocking} & The attacker cannot satisfy the predicate. & \texttt{github.actor == 'trusted-user'}; \texttt{author\_association == 'MEMBER'}\\
\texttt{trusted-action} & The path requires a trusted actor action. & \texttt{github.event.label.name == 'safe'}; \texttt{github.event.review.state == 'approved'} \\
\texttt{passable} & The attacker can satisfy the predicate. & \texttt{contains(github.event.comment.body, '/run')}; \texttt{github.event.issue.body != ''} \\
\texttt{event-filter} & Restricts which events can reach the node. & \texttt{github.event\_name == 'issue\_comment'}; \texttt{github.event\_name == 'pull\_request'} \\
\bottomrule
\end{tabular}
}
\end{table*}

%% file: algorithms/triggerability-analysis.tex
\begin{algorithm}[t]
\caption{Triggerability Analysis}
\label{alg:triggerability}
\small
\KwInput{L-WPG $G$, LLM node $n$, policy registry $\mathcal{P}$}
\KwOutput{$\mathit{Triggerability}(n)$, $\mathit{StepReachability}(n)$}
$C_n^{(0)} \gets \{(e, a) \in G.\mathsf{on} \mid \mathrm{mode}(e, a) \neq \mathit{none}\}$\;
$\mathsf{secretReq}(n) \gets \textsc{ProfileActivation}(n)$\;
$\mathcal{J}(n) \gets \mathrm{toposort}(\mathrm{ancestors}(J_n) \cup \{J_n\})$\;
$C_n^{\mathsf{guard}} \gets \textsc{FilterByGuards}(C_n^{(0)},\; \mathcal{J}(n),\; s_n)$\;
$\mathit{baseline}(n) \gets \textsc{BestMode}(C_n^{\mathsf{guard}})$\;
$C_n^{\mathsf{act}} \gets \textsc{ActivationGate}(C_n^{\mathsf{guard}},\; \mathsf{secretReq}(n))$\;
$C_n \gets \textsc{PolicyGate}(C_n^{\mathsf{act}},\; n,\; \mathcal{P})$\;
$\mathit{mode}(n) \gets \textsc{BestMode}(C_n)$\;
$\mathit{Triggerability}(n) \gets \langle C_n,\; \mathit{baseline}(n),\; \mathit{mode}(n) \rangle$\;
\lIf{$C_n = \emptyset$}{
  \Return $\mathit{Triggerability}(n), \emptyset$
  }
$R(n) \gets \mathrm{ancestors}(J_n) \cup \{J_n\} \cup \mathrm{downstream}(J_n)$\;
\For{\textbf{each} step $s \in R(n)$}{
  $G(s) \gets \mathrm{ancestors}(J_s) \cup \{J_s\}$\;
  $C_{\mathrm{raw}}(s) \gets \textsc{ReachableTriggers}(s,\; G(s))$\;
  $C_{\mathrm{eff}}(s,n) \gets C_{\mathrm{raw}}(s) \cap C_n$\;
  $\mathit{StepReachability}(n)[s] \gets (C_{\mathrm{raw}}(s), C_{\mathrm{eff}}(s,n))$\;
}
\Return $\mathit{Triggerability}(n), \mathit{StepReachability}(n)$\;
\end{algorithm}

%% file: tables/taint-sources.tex
\begin{table}[H]
\centering
\caption{Canonical attacker-controlled taint sources tracked by \tool.}
\label{tab:taint-sources}
\small
\renewcommand{\arraystretch}{1.1}
\setlength{\tabcolsep}{4pt}
\resizebox{\linewidth}{!}{%
\begin{tabular}{ll}
\toprule
\textbf{Category} & \textbf{Canonical Source Expression} \\
\midrule
Issue      & \texttt{github.event.issue.title}              \\
           & \texttt{github.event.issue.body}               \\
Comment    & \texttt{github.event.comment.body}             \\
Pull Request (PR) & \texttt{github.event.pull\_request.title}      \\
           & \texttt{github.event.pull\_request.body}       \\
           & \texttt{github.event.pull\_request.head.ref}   \\
           & \texttt{github.event.pull\_request.head.label} \\
PR review  & \texttt{github.event.review.body}              \\
Discussion & \texttt{github.event.discussion.title}         \\
           & \texttt{github.event.discussion.body}          \\
\bottomrule
\end{tabular}%
}
\end{table}

%% file: algorithms/dataflow-analysis.tex
\begin{algorithm}[!b]
\caption{Dataflow Analysis}
\label{alg:dataflow-analysis}
\small
\KwInput{L-WPG $G$, LLM node $n$, $\mathit{Triggerability}(n)$, $\mathit{StepReachability}(n)$, cache $\mathcal{C}$}
\KwOutput{$\mathit{DataflowView}(n)$}
$R(n) \gets \mathrm{ancestors}(J_n) \cup \{J_n\} \cup \mathrm{downstream}(J_n)$\;
$\mathcal{S}(n) \gets \{s \in \mathrm{steps}(R(n)) \mid C_{\mathrm{eff}}(s,n) \neq \emptyset\}$\;
$\Sigma(s_n) \gets \textsc{Summarize}(s_n,\; \mathit{llm\_target},\; \mathcal{C})$\;
\For{\textbf{each} step $s \in \mathcal{S}(n) \setminus \{s_n\}$ in topo-order}{
  $\Sigma(s) \gets \textsc{Summarize}(s,\; \mathit{ordinary},\; \mathcal{C})$\;
}
$\mathcal{T}$, $\mathcal{L}$, $P_n$, $X_n \gets \emptyset$\;
$W_n$, $R_n^{\mathrm{exec}}$, $R_n^{\mathrm{dec}}$, $R_n^{\mathrm{cont}}$, $A_n \gets \mathit{false}$\;

\For{\textbf{each} step $s \in \mathcal{S}(n)$ in topo-order}{
  Resolve $\Sigma(s).\mathsf{inflows}$ against $\mathcal{T}$, $\mathcal{L}$, workspace contamination, and $C_{\mathrm{eff}}(s,n)$\;
  \If{$s = s_n$}{
    Update $P_n$, $W_n$, and seed $\mathcal{L}$ from the prompt/response channels of $\Sigma(s_n)$\;
  }
  \For{\textbf{each} tainted outflow $o \in \Sigma(s).\mathsf{outflows}$}{
    Update carrier state in $\mathcal{T}$ and $\mathcal{L}$ (outputs, env vars, workspace, artifacts)\;
    Update $X_n$, $R_n^{\mathrm{exec}}$, $R_n^{\mathrm{dec}}$, $R_n^{\mathrm{cont}}$, and $A_n$ from $o$'s sink kind and taint kind\;
  }
}
\Return $\mathit{DataflowView}(n)$\;
\end{algorithm}

%% file: tables/sink-taxonomy.tex
\begin{table}[H]
\centering
\caption{Sink taxonomy used by \tool's dataflow analysis.
  Propagation sinks carry taint to successor steps.
  LLM virtual sinks are internal analysis constructs that demarcate the \ac{llm} interaction boundary.}
\label{tab:sink-taxonomy}
\small
\renewcommand{\arraystretch}{1.1}
\setlength{\tabcolsep}{5pt}
\resizebox{\linewidth}{!}{%
\begin{tabular}{ll}
\toprule
\textbf{Category} & \textbf{Sink Types} \\
\midrule
Execution
  & \texttt{\{yaml\_splice,process,github\_script\}\_exec}, \texttt{shell\_eval} \\
Decision
  & \texttt{\{spam,pr\_review,check\_status,label,deploy\}\_gate} \\
Content
  & \texttt{github\_\{comment,pr\_body,file\_write,release\}}, \\
  & \texttt{workspace\_file\_write}, \texttt{artifact\_upload} \\
Propagation
  & \texttt{\{step,job\}\_output}, \texttt{env\_var}, \\
  & \texttt{workspace\_file\_write}, \texttt{artifact\_upload} \\
LLM virtual
  & \texttt{llm\_\{prompt,response\}}, \texttt{agent\_control} \\
\bottomrule
\end{tabular}%
}
\end{table}

%% file: reference.bib
@article{gallaba2018use,
  title={Use and misuse of continuous integration features: An empirical study of projects that (mis) use Travis CI},
  author={Gallaba, Keheliya and McIntosh, Shane},
  journal={IEEE Transactions on Software Engineering},
  volume={46},
  number={1},
  pages={33--50},
  year={2018},
  publisher={IEEE}
}

@inproceedings{vassallo2020configuration,
  title={Configuration smells in continuous delivery pipelines: a linter and a six-month study on GitLab},
  author={Vassallo, Carmine and Proksch, Sebastian and Jancso, Anna and Gall, Harald C and Di Penta, Massimiliano},
  booktitle={Proceedings of the 28th ACM joint meeting on european software engineering conference and symposium on the foundations of software engineering},
  pages={327--337},
  year={2020}
}

@inproceedings{koishybayev2022characterizing,
  title={Characterizing the security of github $\{$CI$\}$ workflows},
  author={Koishybayev, Igibek and Nahapetyan, Aleksandr and Zachariah, Raima and Muralee, Siddharth and Reaves, Bradley and Kapravelos, Alexandros and Machiry, Aravind},
  booktitle={31st USENIX Security Symposium (USENIX Security 22)},
  pages={2747--2763},
  year={2022}
}

@inproceedings{li2022robbery,
  title={Robbery on devops: Understanding and mitigating illicit cryptomining on continuous integration service platforms},
  author={Li, Zhi and Liu, Weijie and Chen, Hongbo and Wang, XiaoFeng and Liao, Xiaojing and Xing, Luyi and Zha, Mingming and Jin, Hai and Zou, Deqing},
  booktitle={2022 IEEE Symposium on Security and Privacy (SP)},
  pages={2397--2412},
  year={2022},
  organization={IEEE}
}

@inproceedings{gu2023continuous,
  title={Continuous intrusion: Characterizing the security of continuous integration services},
  author={Gu, Yacong and Ying, Lingyun and Chai, Huajun and Qiao, Chu and Duan, Haixin and Gao, Xing},
  booktitle={2023 IEEE Symposium on Security and Privacy (SP)},
  pages={1561--1577},
  year={2023},
  organization={IEEE}
}

@article{pan2023ambush,
  title={Ambush from all sides: Understanding security threats in open-source software ci/cd pipelines},
  author={Pan, Ziyue and Shen, Wenbo and Wang, Xingkai and Yang, Yutian and Chang, Rui and Liu, Yao and Liu, Chengwei and Liu, Yang and Ren, Kui},
  journal={IEEE Transactions on Dependable and Secure Computing},
  volume={21},
  number={1},
  pages={403--418},
  year={2023},
  publisher={IEEE}
}

@inproceedings{muralee2023argus,
  title={$\{$ARGUS$\}$: A Framework for Staged Static Taint Analysis of $\{$GitHub$\}$ Workflows and Actions},
  author={Muralee, Siddharth and Koishybayev, Igibek and Nahapetyan, Aleksandr and Tystahl, Greg and Reaves, Brad and Bianchi, Antonio and Enck, William and Kapravelos, Alexandros and Machiry, Aravind},
  booktitle={32nd USENIX Security Symposium (USENIX Security 23)},
  pages={6983--7000},
  year={2023}
}

@inproceedings{kimsok2026sok,
  title={SoK: Attack and Defense Landscape of Agentic AI Systems},
  author={Kim, Juhee and Guo, Wenbo and Song, Dawn},
  booktitle={35nd USENIX Security Symposium (USENIX Security 26)},
  year={2026}
}

@inproceedings{li2024toward,
  title={Toward Understanding the Security of Plugins in Continuous Integration Services},
  author={Li, Xiaofan and Gu, Yacong and Qiao, Chu and Zhang, Zhenkai and Liu, Daiping and Ying, Lingyun and Duan, Haixin and Gao, Xing},
  booktitle={Proceedings of the 2024 on ACM SIGSAC Conference on Computer and Communications Security},
  pages={482--496},
  year={2024}
}

@inproceedings{gu2024more,
  title={More haste, less speed: Cache related security threats in continuous integration services},
  author={Gu, Yacong and Ying, Lingyun and Chai, Huajun and Pu, Yingyuan and Duan, Haixin and Gao, Xing},
  booktitle={2024 IEEE Symposium on Security and Privacy (SP)},
  pages={1179--1197},
  year={2024},
  organization={IEEE}
}

@article{sun2025does,
  title={Does ai code review lead to code changes? a case study of github actions},
  author={Sun, Kexin and Kuang, Hongyu and Baltes, Sebastian and Zhou, Xin and Zhang, He and Ma, Xiaoxing and Rong, Guoping and Shao, Dong and Treude, Christoph},
  journal={arXiv preprint arXiv:2508.18771},
  year={2025}
}

@inproceedings{tystahl2026cosseter,
  title={COSSETER: GitHub Actions Permission Reduction Using Demand-Driven Static Analysis},
  author={Tystahl, Greg and Ghebremichael, Jonah and Muralee, Siddharth and Cherupattamoolayil, Sourag and Bianchi, Antonio and Machiry, Aravind and Kapravelos, Alexandros and Enck, William},
  booktitle={2026 IEEE Symposium on Security and Privacy (SP)},
  year={2026}
}

@inproceedings{liu2024demystifying,
  title={Demystifying rce vulnerabilities in llm-integrated apps},
  author={Liu, Tong and Deng, Zizhuang and Meng, Guozhu and Li, Yuekang and Chen, Kai},
  booktitle={Proceedings of the 2024 on ACM SIGSAC Conference on Computer and Communications Security},
  pages={1716--1730},
  year={2024}
}

@inproceedings{liu2025make,
  title={Make agent defeat agent: Automatic detection of $\{$Taint-Style$\}$ vulnerabilities in $\{$LLM-based$\}$ agents},
  author={Liu, Fengyu and Zhang, Yuan and Luo, Jiaqi and Dai, Jiarun and Chen, Tian and Yuan, Letian and Yu, Zhengmin and Shi, Youkun and Li, Ke and Zhou, Chengyuan and others},
  booktitle={34th USENIX Security Symposium (USENIX Security 25)},
  pages={3767--3786},
  year={2025}
}

@inproceedings{he2026taintp2x,
  title={TaintP2X: Detecting Taint-Style Prompt-to-Anything Injection Vulnerabilities in LLM-Integrated Applications},
  author={He, Junjie and Wang, Shenao and Zhao, Yanjie and Hou, Xinyi and Liu, Zhao and Zou, Quanchen and Wang, Haoyu},
  booktitle={2026 IEEE/ACM 48th International Conference on Software Engineering (ICSE)},
  year={2026}
}

@inproceedings{greshake2023not,
  title={Not what you've signed up for: Compromising real-world llm-integrated applications with indirect prompt injection},
  author={Greshake, Kai and Abdelnabi, Sahar and Mishra, Shailesh and Endres, Christoph and Holz, Thorsten and Fritz, Mario},
  booktitle={Proceedings of the 16th ACM workshop on artificial intelligence and security},
  pages={79--90},
  year={2023}
}

@article{hines2024defending,
  title={Defending against indirect prompt injection attacks with spotlighting},
  author={Hines, Keegan and Lopez, Gary and Hall, Matthew and Zarfati, Federico and Zunger, Yonatan and Kiciman, Emre},
  journal={arXiv preprint arXiv:2403.14720},
  year={2024}
}

@inproceedings{chen2025struq,
  title={$\{$StruQ$\}$: Defending against prompt injection with structured queries},
  author={Chen, Sizhe and Piet, Julien and Sitawarin, Chawin and Wagner, David},
  booktitle={34th USENIX Security Symposium (USENIX Security 25)},
  pages={2383--2400},
  year={2025}
}

@inproceedings{zhong2026attention,
  title={Attention is All You Need to Defend Against Indirect Prompt Injection Attacks in LLMs},
  author={Zhong, Yinan and Miao, Qianhao and Chen, Yanjiao and Deng, Jiangyi and Cheng, Yushi and Xu, Wenyuan},
  booktitle={Proceedings of the 33rd Network and Distributed System Security (NDSS) Symposium},
  year={2026}
}

@inproceedings{shi2024optimization,
  title={Optimization-based prompt injection attack to llm-as-a-judge},
  author={Shi, Jiawen and Yuan, Zenghui and Liu, Yinuo and Huang, Yue and Zhou, Pan and Sun, Lichao and Gong, Neil Zhenqiang},
  booktitle={Proceedings of the 2024 on ACM SIGSAC Conference on Computer and Communications Security},
  pages={660--674},
  year={2024}
}

@inproceedings{shi2026prompt,
  title={Prompt injection attack to tool selection in llm agents},
  author={Shi, Jiawen and Yuan, Zenghui and Tie, Guiyao and Zhou, Pan and Gong, Neil Zhenqiang and Sun, Lichao},
  booktitle={Proceedings of the 33rd Network and Distributed System Security (NDSS) Symposium},
  year={2026}
}

@inproceedings{chen2025defending,
  title={Defending against prompt injection with a few defensivetokens},
  author={Chen, Sizhe and Wang, Yizhu and Carlini, Nicholas and Sitawarin, Chawin and Wagner, David},
  booktitle={Proceedings of the 18th ACM Workshop on Artificial Intelligence and Security},
  pages={242--252},
  year={2025}
}

@article{liu2025your,
  title={" Your AI, My Shell": Demystifying Prompt Injection Attacks on Agentic AI Coding Editors},
  author={Liu, Yue and Zhao, Yanjie and Lyu, Yunbo and Zhang, Ting and Wang, Haoyu and Lo, David},
  journal={arXiv preprint arXiv:2509.22040},
  year={2025}
}

@inproceedings{liu2024formalizing,
  title={Formalizing and benchmarking prompt injection attacks and defenses},
  author={Liu, Yupei and Jia, Yuqi and Geng, Runpeng and Jia, Jinyuan and Gong, Neil Zhenqiang},
  booktitle={33rd USENIX Security Symposium (USENIX Security 24)},
  pages={1831--1847},
  year={2024}
}

@misc{daelman2025promptpwnd,
  author = {Rein Daelman},
  title = {PromptPwnd: Prompt Injection Vulnerabilities in GitHub Actions Using AI Agents},
  howpublished = {\url{https://www.aikido.dev/blog/promptpwnd-github-actions-ai-agents}},
  year = {2025},
  month = {12},
  day = {4}
}

@misc{githubContinuousIntegration,
	author = {GitHub},
	title = {{C}ontinuous integration - {G}it{H}ub {D}ocs --- docs.github.com},
	howpublished = {\url{https://docs.github.com/en/actions/get-started/continuous-integration}},
	year = {2026}
}

@misc{githubWorkflowSyntax,
	author = {GitHub},
	title = {{W}orkflow syntax for {G}it{H}ub {A}ctions - {G}it{H}ub {D}ocs --- docs.github.com},
	howpublished = {\url{https://docs.github.com/en/actions/reference/workflows-and-actions/workflow-syntax}},
	year = {2026}
}

@misc{githubGitHubActionsaiinference,
	author = {GitHub},
	title = {{G}it{H}ub - actions/ai-inference: {A}n action for calling {A}{I} models with {G}it{H}ub {M}odels --- github.com},
	howpublished = {\url{https://github.com/actions/ai-inference}},
	year = {2026}
}

@misc{githubGitHubJohnsonRannomorespam,
	author = {GitHub},
	title = {{G}it{H}ub - {J}ohnson{R}an/nomore-spam --- github.com},
	howpublished = {\url{https://github.com/{J}ohnson{R}an/nomore-spam}},
	year = {2026}
}

@misc{githubGitHubOpenaicodexaction,
	author = {OpenAI},
	title = {{G}it{H}ub - openai/codex-action --- github.com},
	howpublished = {\url{https://github.com/openai/codex-action}},
	year = {2026}
}

@misc{googleGeminiFlash,
	author = {Google},
	title = {{G}emini 3 {F}lash {P}review  |  {G}emini {A}{P}{I}  |  {G}oogle {A}{I} for {D}evelopers --- ai.google.dev},
	howpublished = {\url{https://ai.google.dev/gemini-api/docs/models/gemini-3-flash-preview}},
	year = {2026}
}

@misc{githubGitHubAnthropicsclaudecodeaction,
	author = {GitHub},
	title = {{G}it{H}ub - anthropics/claude-code-action --- github.com},
	howpublished = {\url{https://github.com/anthropics/claude-code-action}},
	year = {2026}
}

@misc{microsoftContentFiltering,
	author = {ssalgadodev},
	title = {{C}ontent filtering for {M}icrosoft {F}oundry {M}odels (classic) - {M}icrosoft {F}oundry (classic) portal --- learn.microsoft.com},
	howpublished = {\url{https://learn.microsoft.com/en-us/azure/foundry-classic/foundry-models/concepts/content-filter}},
	year = {2026}
}

@article{zheng2023judging,
  title={Judging llm-as-a-judge with mt-bench and chatbot arena},
  author={Zheng, Lianmin and Chiang, Wei-Lin and Sheng, Ying and Zhuang, Siyuan and Wu, Zhanghao and Zhuang, Yonghao and Lin, Zi and Li, Zhuohan and Li, Dacheng and Xing, Eric and others},
  journal={Advances in neural information processing systems},
  volume={36},
  pages={46595--46623},
  year={2023}
}

@misc{gharchiveArchive,
	author = {GHArchive},
	title = {{G}{H} {A}rchive --- gharchive.org},
	howpublished = {\url{https://www.gharchive.org/}},
	year = {2026}
}

@misc{githubGitHubMarketplace,
	author = {GitHub},
	title = {{G}it{H}ub {M}arketplace: tools to improve your workflow},
	howpublished = {\url{https://github.com/marketplace}},
	year = {2026}
}

@misc{githubReuseWorkflows,
	author = {GitHub},
	title = {{R}euse workflows - {G}it{H}ub {D}ocs --- docs.github.com},
	howpublished = {\url{https://docs.github.com/en/actions/how-tos/reuse-automations/reuse-workflows}},
	year = {2026}
}

@misc{githubDisallowAccess,
	author = {GitHub},
	title = {{D}isallow access to secrets for pull\_request trigger · community · {D}iscussion \#180109 --- github.com},
	howpublished = {\url{https://github.com/orgs/community/discussions/180109}},
	year = {2026}
}

@misc{githubToolkitpackagesgithubMain,
	author = {GitHub},
	title = {toolkit/packages/github at main · actions/toolkit --- github.com},
	howpublished = {\url{https://github.com/actions/toolkit/tree/main/packages/github}},
	year = {2026}
}

@misc{githubToolkitpackagescoreMain,
	author = {GitHub},
	title = {toolkit/packages/core at main · actions/toolkit --- github.com},
	howpublished = {\url{https://github.com/actions/toolkit/tree/main/packages/core}},
	year = {2026}
}

@misc{githubGitHubgh,
	author = {GitHub},
	title = {{G}it{H}ub {C}{L}{I} --- cli.github.com},
	howpublished = {\url{https://cli.github.com/}},
	year = {2026}
}

@misc{githubOctokit,
	author = {GitHub},
	title = {{O}ctokit --- github.com},
	howpublished = {\url{https://github.com/octokit}},
	year = {2026}
}
